\documentclass[12pt,a4paper]{article}
\usepackage{amsmath,amssymb}
\usepackage{geometry}
\usepackage{cite}
\usepackage{mathrsfs}
\usepackage{axodraw}
\usepackage{subfigure}
\usepackage{enumerate}
\usepackage{hhline}
\usepackage{multirow}
\usepackage{bigdelim}
\usepackage{bbm}
\usepackage{wrapfig}
\usepackage{sidecap}
\usepackage{caption}
\usepackage[T1]{fontenc}

\renewcommand{\slash}[2][4]{\ensuremath{\rlap{\raisebox{1pt}{$\mspace{#1mu}/$}}#2}}
\newlength{\overlinelength}
\newlength{\textlength}
\newcommand{\ol}[2][.625]{%
   \settowidth{\textlength}{$#2$}%
   \setlength{\overlinelength}{3pt}%
   \addtolength{\overlinelength}{0.4\textlength}%
   \makebox[\textlength][s]{$#2$}%
   \hspace{-.5\textlength}\hspace{-\overlinelength}\hspace{#1\overlinelength}
   \overline{%
      \makebox[\overlinelength][s]{%
         \vphantom{$#2$}
      }
   }
   \hspace{-#1\overlinelength}\hspace{.5\textlength}
}
\def\clap#1{\hbox to 0pt{\hss#1\hss}}

\DeclareMathOperator{\tr}{tr}
\DeclareMathOperator{\diag}{diag}

\newcommand{\cQb}{\ensuremath{\ol{\mathcal{Q}}}}
\newcommand{\Qb}{\ensuremath{\ol{Q}}}
\newcommand{\Db}{\ensuremath{\ol{D}}}
\newcommand{\Fb}{\ensuremath{\ol{F}}}
\newcommand{\Ab}{\ensuremath{\ol{A}}}
\newcommand{\thetab}{\ensuremath{\bar{\theta}}}
\newcommand{\phib}{\ensuremath{\ol[.8]{\phi}}}
\newcommand{\psib}{\ensuremath{\bar{\psi}}}
\newcommand{\sigmab}{\ensuremath{\bar{\sigma}}}
\newcommand{\lambdab}{\ensuremath{\bar{\lambda}}}
\newcommand{\chib}{\ensuremath{\bar{\chi}}}
\newcommand{\etab}{\ensuremath{\bar{\eta}}}
\newcommand{\partialb}{\ensuremath{\bar{\partial}}}
\newcommand{\Psib}{\ensuremath{\bar{\Psi}}}

\newcommand{\mb}{\ensuremath{\ol{m}}}
\newcommand{\varphib}{\ensuremath{\bar{\varphi}}}
\newcommand{\alphab}{\ensuremath{\bar{\alpha}}}
\newcommand{\alphad}{{\ensuremath{\dot{\alpha}}}}
\newcommand{\betad}{{\ensuremath{\dot{\beta}}}}
\newcommand{\Qt}{\ensuremath{\widetilde{Q}}}
\newcommand{\Ft}{\ensuremath{\widetilde{F}}}
\newcommand{\varphit}{\ensuremath{\tilde{\varphi}}}
\renewcommand{\i}{\ensuremath{\text{i}}}
\newcommand{\2}{\ensuremath{\sqrt{2}\,}}
\renewcommand{\d}{\ensuremath{\text{d}}}
\renewcommand{\L}{\ensuremath{\mathscr{L}}}
\newcommand{\dslash}{\slash[2]{\partial}}
\newcommand{\Dslash}{\slash{D}}
\DeclareTextSymbol{\textL}{T1}{138}
\DeclareTextSymbolDefault{\textL}{T1}
\begin{document}
  \hspace*{\fill} DESY-04-248\\

  \begin{center}
    {\LARGE From Super-Yang-Mills to QCD}\\
    \vspace{10mm}
    {\Large Christoph L\"{u}deling\large\footnote{\ttfamily christoph.luedeling@desy.de}}\\[3mm]
    \slshape Deutsches Elektronensynchrotron DESY\\
    \slshape Notkestr. 85\\
    \slshape 22603 Hamburg\\
    \slshape Germany
  \end{center}
  
  \begin{abstract}
    This article contains lecture notes of M.~Shifman from the Saalburg Summer School
    2004. The topic is supersymmetric Yang-Mills theory, in particular the gluino condensate in
    pure SUSY gluodynamics.
  \end{abstract}
  
  \section{Introduction}
    These are Lecture Notes from the Saalburg Summer School 2004 Lecture of M.~Shifman, William
    I. Fine Theoretical Physics Institute, University of Minnesota, Minneapolis, MN 55455, USA. They
    deal with supersymmetric gauge theories, in particular with the calculation of the gluino
    condensate. A more in-depth treatment can be found in the review \cite{shifmanvainshtein}
    and the book \cite{shifmanbook}. 
    A concise and very clearly written introduction to supersymmetry
    can be found in the book \cite{bailin} which I also recommend.
    There are various older papers on the subject, see
    e.g. \cite{shifmanvainshteinold,affleckdineseibergetal}. 

  \section{SUSY Preliminaries}
    \subsection{SUSY Algebra}
      In contrast to the generators of the Lorentz group, the generators of supersymmetry
      transformations (``supercharges'') are spinors and obey anticommutation relations. In
      this lectures we only consider $\mathcal{N}=1$ supersymmetry in four spacetime
      dimensions. Then the SUSY generators are a left-handed Weyl spinor $\mathcal{Q}_\alpha$,
      $\alpha=1,2$, and its Hermitean conjugate $\cQb_{\alphad}$, $\alphad=1,2$ which  satisfy
      the (anti)commutation relations \cite{golfand}
      \begin{subequations}
      \begin{align}
           \left\{\mathcal{Q}_\alpha,\mathcal{Q}_\beta\vphantom{\cQb}\right\}& =
          \left\{\cQb_{\alphad},\cQb_{\betad}\right\}=0\\
          \left\{\mathcal{Q}_\alpha,\cQb_{\alphad}\right\}& =
            2\sigma^m_{\alpha\alphad} \mathcal{P}_m \label{susyalgebra-QQb}\\
          \left[\mathcal{Q}_\alpha,\mathcal{P}_m\vphantom{\cQb}\right]&
          =\left[\cQb_{\alphad},\mathcal{P}_m\right]=0 \label{susyalgebra-QP}
      \end{align}
      \end{subequations}
      From Eq. (\ref{susyalgebra-QQb}), wee see that the mass dimension of $\mathcal{Q}$ is
      $\left[\mathcal{Q}_\alpha\right]=\tfrac{1}{2}\left[\mathcal{P}_m\right]=\tfrac{1}{2}$.
      The last line tells us that $\left[\mathcal{Q}_\alpha,\mathcal{P}^2\right]=0$, which
      means that supersymmetry transformations do not change the mass of a state.
      
      Poincar\'{e} and supersymmetry basically exhaust the possible symmetries of the S matrix
      of a sensible QFT, apart from a compact internal symmetry\footnote{This is roughly the
        Haag--\textL opuszanski--Sohnius theorem, see \cite{haag}.}. The part of the internal
      symetry which does not commute with the generators of supersymmetry transformations is
      called $R$ symmetry, which in the $\mathcal{N}=1$ case is at most a $\sf U(1)$ rotating
      the supercharges. 

    \subsection{Superspace}
      Just as the momentum operator is realised as translations $-\i\partial_m$ in Minkowski
      space, the supersymmetry generators can be represented by differential operators
      $Q_\alpha$, $\Qb_{\alphad}$ on a larger space which is (of course) called superspace (it
      was introduced in \cite{salam}).  In 
      addition to the usual Minkowski coordinates $x^m$, this space contains two additional
      Grassmann valued (i.e.\ anticommuting) Weyl spinors $\theta^\alpha$,
      $\thetab_{\alphad}$. Points in this space are thus labeled by coordinates
      $z^M=(x^m,\theta^\alpha,\thetab_{\alphad})$. (Our conventions regarding spinors and
      $\sigma$-matrices are collected in the appendix.) Note that the $\theta$'s have mass dimension
      $\left[\theta\right]=\left[\thetab\mspace{2mu}\right]=-\tfrac{1}{2}$. If an $R$ symmetry
      is present, the $\theta$'s are rotated as well. The supersymmetry generators are then
      represented by the operators
      \begin{align}
        \begin{split}
          Q_\alpha&= \partial_\alpha -\i\sigma^m_{\alpha\alphad}\thetab^{\alphad}\partial_m\,,\\
          \Qb_{\alphad} &=-\partialb_{\alphad}
          +\i\theta^\alpha\sigma^m_{\alpha\alphad}\partial_m.
        \end{split}
      \end{align}
      
      Other important operators are the supersymmetry-covariant derivatives
      \begin{align}
        \begin{split}
          D_\alpha&= \partial_\alpha +\i\sigma^m_{\alpha\alphad}\thetab^{\alphad}\partial_m\\
          \Db_{\alphad} &=-\partialb_{\alphad}
          -\i\theta^\alpha\sigma^m_{\alpha\alphad}\partial_m
        \end{split}
      \end{align}
      which fulfill
      \begin{gather*}
        \left\{D_\alpha,\Db_{\alphad}\right\}=-2\i\sigma^m_{\alpha\alphad}\partial_m\,, \qquad
        \left\{D_\alpha,D_\beta\right\}=\left\{\Db_{\alphad},\Db_{\betad}\right\}=0\,,\\
        \left\{D_\alpha,Q_\beta\right\}= \left\{D_\alpha,\Qb_{\betad}\right\}=
        \left\{\Db_{\alphad},Q_\beta\right\}= \left\{\Db_{\alphad},\Qb_{\betad}\right\}=0.
      \end{gather*}

    \subsection{Superfields}
      Superfields (which were introduced in the 
      same work of Salam and Strathdee) 
      are functions of the superspace coordinates which are defined by their expansion in the
      anticommuting coordinates. Since $\theta^3=\thetab^3=0$, the expansion can contain only
      terms with at most two $\theta$'s and two $\thetab$'s:
      \begin{align}
        \begin{split}\label{generalsuperfield}
          F(x,\theta,\thetab)&=b+\theta\chi +\thetab\phib +\theta^2 m +\thetab^2 n
          +\theta\sigma^m\thetab a_m +\theta^2\thetab\psib +\thetab^2\theta\lambda
          +\theta^2\thetab^2 d,
        \end{split}
      \end{align}
      where the component fields $b$, $m$, $n$ and $d$ are scalars, $a_m$ is a vector and
      $\chi$, $\lambda$, $\psib$ and $\phib$ are left- or right-handed spinors, all functions
      of $x^m$. Moreover,
      $$
        \theta^2 \equiv \theta^\alpha\theta_\alpha\,,\qquad \bar\theta^2 \equiv
        \theta_{\dot\alpha}\theta^{\dot\alpha}\,. 
      $$
      One may or may not impose the condition of reality, $F^\dagger = F$. The superfield $F$
      is usually referred to as the vector superfield. The reason will become clear
      shortly. The component  fields in $F$ form a representation of the supersymmetry
      algebra. They will be involved in construction of the gauge sector. For constructing
      supersymmetric Yang-Mills theories with matter, we will also use shorter superfields as
      generalisations of matter fermions. The matter fields will be components of the
      (irreducible) chiral multiplet $\phi$, defined as 
      \begin{subequations}
      \begin{align}
        \Db_{\alphad} \phi &=0 \,.
        \label{chiralcondition}
      \end{align}
      The chiral superfield  can be regarded as a function defined  not on the full superspace,
      but, rather,  on the left-handed subspace parametrised by the coordinates
      $x_{\scriptscriptstyle L}^m= x^m+\i\theta\sigma^m\thetab$ and $\theta$, with no explicit
      dependence on $\thetab$, since in this basis, $\Db_{\alphad}=-\partialb_{\alphad}$, and
      thus Eq. (\ref{chiralcondition}) means  that $\phi$ does not depend on $\thetab$
      explicitly. The possibility of introducing such superfield is due to the fact that the
      supertransformations of  $x_{\scriptscriptstyle L}$ and $\theta$ are closed, namely,
      \begin{align*}
        \delta_\eta x_{\scriptscriptstyle L}^m &= 2\i\theta\sigma^m\etab & \delta_\eta
        \theta^\alpha &= \eta^\alpha,
      \end{align*}
      where $\eta$ is the (Weyl spinor valued) transformation parameter.
      
      The component expansion of the chiral superfield is particularly simple. The expansion in
      terms of the usual coordinates can be obtained by the Taylor expansion in the Grassmann
      variables, 
      \begin{align}
        \begin{split}\label{chiralsf}
          \phi&= A(x_{\scriptscriptstyle L}) +\2\theta\chi(x_{\scriptscriptstyle L}) +\theta^2
          F(x_{\scriptscriptstyle L}) \\
            &=A(x) +\i\theta\sigma^m\thetab\partial_m A(x) +\tfrac{1}{4}\theta^2\thetab^2\Box
            A(x)\\
            &\quad +\2\theta\chi(x) -\tfrac{\i}{\2}\theta^2\partial_m\chi(x)\sigma^m\thetab
            +\theta^2 F(x)  \,.
        \end{split}
      \end{align}
      
      Analogously, antichiral fiels $\phib$ are defined as 
      \begin{align}
        D_\alpha \phib&=0,
      \end{align}
      and their expansion is best expressed in terms of $x_{\scriptscriptstyle R}^m= x^m-
      \i\theta\sigma^m\thetab$ and $\thetab$: 
      \begin{align}
        \begin{split}\label{antichiralsf}
          \phib&= \Ab(x_{\scriptscriptstyle R}) +\2\thetab\chib(x_{\scriptscriptstyle R})
          +\theta^2 \Fb(x_{\scriptscriptstyle R}) \\
            &=\Ab(x) -\i\theta\sigma^m\thetab\partial_m \Ab(x) +\tfrac{1}{4}\theta^2\thetab^2\Box
            \Ab(x)\\
            &\quad +\2\thetab\chib(x) +\tfrac{\i}{\2}\thetab^2\theta\sigma^m\partial_m\chib(x)
            +\theta^2 \Fb(x) \,.
        \end{split}
      \end{align}

      Conjugates of chiral fields are antichiral, and a field that is both chiral and
      antichiral is constant. Products of chiral fields are again chiral, a product of a chiral
      and an antichiral field is neither chiral nor antichiral.
      \end{subequations}

      If we want the scalar $A$ to have canonical mass dimension $\left[A\right]=1$, we must
      assign to $\phi$ the dimension $\left[\phi\right]=1$. Since
      $\left[\theta\right]=-\tfrac{1}{2}$, the spinor field $\chi$ also has the correct
      dimension. The field $F$, however, has mass dimension two. This signifies that it is an
      auxiliary field, appearing in the Lagrangean without derivatives, and that it can be
      eliminated by its equations of motion.

      The physical components of the chiral superfields are scalar or spinor. To incorporate
      the gauge field we need to turn to Eq. (\ref{generalsuperfield}). The vector superfield
      $V$, which is a generalisation of the gauge vector fields in non-supersymmetric theories,
      contains a vector field, a Weyl fermion and its conjugate (the gaugino, equivalent to a
      Majorana fermion) and a few other fields which can be eliminated by a supergauge
      transformation or are auxiliary. The reality condition 
      \begin{subequations}
      \begin{align}
        V^\dagger &=V
      \end{align}
      is implied.  The component expansion of the vector superfield is given in
      Eq. (\ref{generalsuperfield}). Supergauge
      transformations acting on this field are inhomogeneous. They involve  chiral superfields
      as their parameters, ina way such that lower components of $V$ can be gauged away. This
      is the famous Wess--Zumino gauge which  leaves us with the following decomposition: 
      \begin{align}
        V&=-\theta\sigma^m\thetab v_m +\i\theta^2\thetab\lambdab -\i\thetab^2\theta\lambda
        +\tfrac{1}{2}\theta^2\thetab^2 D.
      \end{align}
      \end{subequations}
      The vector field $v_m$ still has the usual gauge symmetry $v_m\to v_m+\partial_m\beta$ 
      (in the Abelian case). The fields $\lambda$ and $D$ are gauge invariant in the Abelian
      case. In the non-Abelian case they transform homogeneously. The Wess--Zumino gauge is the
      one most frequently used in dealing with supersymmetric gauge theories. 

      On the other hand, if $V$ becomes massive by some kind of the Higgs effect, the most
      convenient gauge is unitary, in which the lower components are no longer   gauged away;
      rather,  they represent additional degrees of freedom, a real scalar $C$ and another Weyl
      fermion $\chi$, as we will see later. 

      The vector superfield contains an auxiliary field as well, the real scalar $D$. Just as
      the $F$ field in the chiral superfield, it has mass dimension two (since we want $v_m$ to
      have canonical dimension). The auxiliary fields are needed to make the algebra close
      off-shell and will be useful when we consider the scalar potential of supersymmetric
      Yang-Mills theories.

    \subsection{Lagrangeans}
      Finally, we need a way to write down invariant actions from superfields. This can be
      achived by constructing Lagrangeans which vary by a total derivative under
      supersymmetry transformations. The highest (auxiliary) components of chiral and real
      superfields have exactly this property. They can be projected out by the rules of
      intergration over the Grassmann variables (Berezin integral\,\footnote{Grassmann variable
        calculus, with applications in quantum field theory, was developed by Felix Berezin in
        a series of papers which were summarized in the book \cite{berezin}. The book was
        published in Russian in 1965, and was translated immediately.}), which for one
      anticommuting variable $\theta$ are\footnote{This completely fixes the integration, since
        any function $f(\theta)$ can be expanded as \mbox{$f(\theta)=a+b\theta$}. This
        definition furthermore is unique up to a factor if we require translational invariance
        of the 
        integral.} 
      \begin{align}
        \int\!\d\theta\, 1 &=0& \int\!\d\theta\,\theta&=1.
      \end{align}
      This means that $\left[\d\theta\right]=-\left[\theta\right]$. Note also that integration
      and differentiation give the same result for the Grassmann variables, so a
      $\d^4\theta$-integration over a chiral field vanishes (actually, it is a total
      derivative, so its spacetime integral vanishes and it does not contribute to the action). 

      For superspace with two Grassmann coordinates, we define
      \begin{align}
        \d^2\theta &= -\tfrac{1}{4}\varepsilon_{\alpha\beta}\d\theta^\alpha\d\theta^\beta \,,
        &
        \d^2\thetab &= -\tfrac{1}{4}\varepsilon^{\alphad\betad}\d\thetab_{\alphad}
        \d\thetab_{\betad}\,,  & 
        \d^4\theta&=\d^2\theta\,\d^2\thetab,
      \end{align}
      with coefficients chosen such that  integration over the superspace coordinates gives
      \begin{align*}
        \int\!\d^2\theta \,\theta^2 &=\int\!\d^2\thetab\,\thetab^2=\int\!\d^4\theta\,
        \theta^2\thetab^2=1.
      \end{align*}

      So, the integral over the Grassmann coordinates projects out the $F$-components of chiral
      superfields and the $D$-components of real superfields. We can for example consider a
      theory with just a chiral superfield $\phi$, i.e.\ a scalar and a spinor. For the
      Lagrangean we need a kinetic term and, maybe, mass or interaction terms. The kinetic term
      is provided just by 
      \begin{subequations}
      \begin{align}
        \L_{\text{kin}}&=\int\d^4\theta\,\phib\phi.
      \end{align}
      The surviving terms need to have two $\theta$'s and two $\thetab$'s, so we have to
      collect the possible combinations from Eqs.\ (\ref{chiralsf}) and
      (\ref{antichiralsf}). This gives 
      \begin{align}\label{simplekaehlerpot}
        \begin{split}
          \L_\text{kin}&=\tfrac{1}{4}\left(\Ab\Box A +\Box \Ab A\right)
          -\tfrac{1}{2}\partial_m\Ab\,\partial^m A +
          \tfrac{\i}{2}\left(\partial_m\chib\sigmab^m\chi -\chib\sigmab^m\partial_m\chi\right)
          +\Fb F \\
          &=-\partial_m \Ab\partial^m A -\i\chib\sigmab^m\partial_m\chi +\Fb F
        \end{split}
      \end{align}
      \end{subequations}
      and we see that indeed $F$ appears without derivatives. In going from the first to the
      second line we have performed an integration by parts and dropped the total derivative
      terms. 

      Mass and interaction terms appear as purely chiral combinations of fields in the so-called
      superpotential,
      \begin{subequations}
      \begin{align}\label{simplesuperpot}
        \L_\text{SP}&=\int\d^2\theta\, \Big( \underbrace{\tfrac{1}{2}m\phi^2
          +\tfrac{1}{3}\lambda\phi^3}_{\textstyle :=W(\phi)} \Big)
        +\text{H.c.},
      \end{align}
      where $m$ and $\lambda$ are in general complex coefficients. In components, this gives 
      \begin{align}
        \L_\text{SP}&= m F A -\tfrac{1}{2} m\chi\chi +\lambda F A^2 - A \chi\chi +\text{H.c.}
      \end{align}
      \end{subequations}
      In Eq. (\ref{simplesuperpot}), a term linear in $\phi$ can be absorbed by a redefinition
      of the field.

      If we now combine the two parts of the Lagrangean to $\L=\L_\text{kin}+\L_\text{SP}$, we
      can eliminate $F$ and $\Fb$ by their equations of motion,
      \begin{align}
        0&=\partial_m \frac{\delta \L}{\delta\left(\partial_m \Fb\right)} =\frac{\delta
          \L}{\delta \Fb} = F +\mb\Ab +\lambdab \Ab^2.
      \end{align}
      This means that the part of the Lagrangean containing $F$ and $\Fb$ reduces to
      \begin{align}
        \L_F&=-\left(\mb\Ab+\lambdab\Ab^2\right)\left(mA+\lambda A^2\right)=
        -\left|\frac{\partial W}{\partial \phi} \right|^2\equiv -V(A),
      \end{align}
      the potential for the scalar field which is always nonnegative.

  \section{Super-Yang-Mills Theory}
    \subsection{Super-QED}
      Super-QED (SQED) is the supersymmetric generalisation of QED, so it must contain an
      electron, and a photon, together with their superpartners, selectron and photino. What
      kinds of interactions can we expect? First of all, there should be the ordinary
      electron-electron-photon vertex from QED \ref{eegamma}. But in  this vertex, we can
      exchange two of the fields for their superpartners, resulting in an
      electron-selectron-photino vertex \ref{ephilambda} and a selectron-selectron-photon
      vertex \ref{phiphigamma}. We will see that there are also two quartic vertices, namely a
      four-selectron vertex \ref{phihochvier} and a selectron-selectron-photon-photon vertex
      \ref{phiphigammagamma}. 
      The form of these vertices is fixed by supersymmetry, as can most convieniently be seen
      in superfields: 

%%%%%%%%%%%%%%%%%%%%%%%%%%%%%%%%%%%%%%%%%%%%%%%%%%%%%%%%%%%%%%%%%%%%%%%%%%%%%%%%%%%%%%%%%%
%%%%%%%%%%%%%%%%%%%%%%%%%%%%% Figure: SQED-Vertices  %%%%%%%%%%%%%%%%%%%%%%%%%%%%%%%%%%%%%

      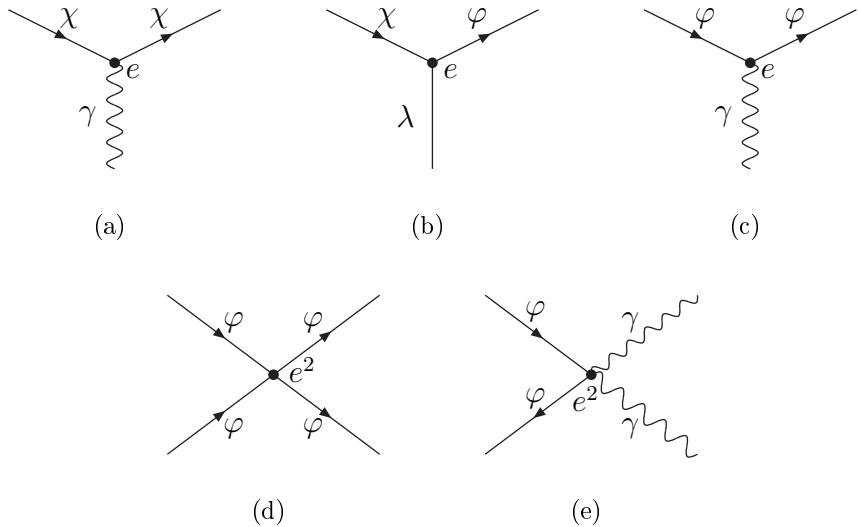
\begin{figure}\label{sqedvertices}
        \begin{center}
          \subfigure[]{\label{eegamma}
            \begin{picture}(80,60)(0,0)
              \Photon(40,40)(40,0){3}{5} \Text(30,20)[]{$\gamma$}
              \ArrowLine(0,60)(40,40) \Text(23,57)[]{$\chi$}
              \ArrowLine(40,40)(80,60) \Text(57,57)[]{$\chi$}
              \Vertex(40,40){2} \Text(47,37)[]{$e$}
            \end{picture}
            }\hspace{1cm}
          \subfigure[]{\label{ephilambda}
            \begin{picture}(80,60)(0,0)
              \Line(40,40)(40,0)  \Text(30,20)[]{$\lambda$}
              \ArrowLine(0,60)(40,40) \Text(23,57)[]{$\chi$}
              \ArrowLine(40,40)(80,60) \Text(57,57)[]{$\varphi$}
              \Vertex(40,40){2} \Text(47,37)[]{$e$}
            \end{picture}
            }\hspace{1cm}
            \subfigure[]{\label{phiphigamma}
            \begin{picture}(80,60)(0,0)
              \Photon(40,40)(40,0){3}{5} \Text(30,20)[]{$\gamma$}
              \ArrowLine(0,60)(40,40) \Text(23,57)[]{$\varphi$}
              \ArrowLine(40,40)(80,60) \Text(57,57)[]{$\varphi$}
              \Vertex(40,40){2} \Text(47,37)[]{$e$}
            \end{picture}
            }

            \subfigure[]{\label{phihochvier}
              \begin{picture}(80,60)(0,0)
                \ArrowLine(0,60)(40,30) \Text(25,50)[]{$\varphi$}
                \ArrowLine(40,30)(80,0)  \Text(55,10)[]{$\varphi$}
                \ArrowLine(0,0)(40,30) \Text(25,10)[]{$\varphi$}
                \ArrowLine(40,30)(80,60) \Text(55,50)[]{$\varphi$}
                \Vertex(40,30){2} \Text(51,32)[]{$e^2$}
              \end{picture}
              }\hspace{1cm}
            \subfigure[]{\label{phiphigammagamma}
              \begin{picture}(80,60)(0,0)
                \ArrowLine(0,60)(40,30) \Text(19,54)[]{$\varphi$}
                \Photon(40,30)(80,0){3}{5}  \Text(55,10)[]{$\gamma$}
                \ArrowLine(40,30)(0,0) \Text(19,21)[]{$\varphi$}
                \Photon(40,30)(80,60){2}{6} \Text(55,50)[]{$\gamma$}
                \Vertex(40,30){2} \Text(38,21)[]{$e^2$}
              \end{picture}
              }
          \caption{Vertices of SQED. Vertex (a) is the ordinary QED
            photon-electron-electron vertex. Exchanging particles and their superpartners gives
            vertices (b) and (c). Additionally, there are the quartic
            scalar and scalar-photon vertices (d) and (e).}
        \end{center}
      \end{figure}
%%%%%%%%%%%%%%%%%%%%%%%%%%%%%%%%%%%%%%%%%%%%%%%%%%%%%%%%%%%%%%%%%%%%%%%%%%%%%%%%%%%%%%%%%%
%%%%%%%%%%%%%%%%%%%%%%%%%%%%%%%%%%%%%%%%%%%%%%%%%%%%%%%%%%%%%%%%%%%%%%%%%%%%%%%%%%%%%%%%%%

      The electron and positron together form a Dirac spinor $\Psi$ which can be decomposed
      into two Weyl spinors of opposite chirality, $\Psi=(\chi,\etab)^T$. Both $\chi$ and $\etab$
      have electric charge $+1$. The complex conjugate field $\Psib=(\eta,\chib)$ also contains
      a left- and a right-handed spinor, both with electric charge $-1$. So we can describe the
      electron by two left handed Weyl spinors $\chi$ and $\eta$ with electric charges~$+1$
      and~$-1$, respectively. The conjugate fields $\chib$ and $\etab$ have opposite
      charges. We  arrange $\chi$ and $\eta$ in two chiral superfields:
      \begin{subequations}
        \begin{align}
          Q   &=\varphi +\theta\chi+\theta^2 F &\text{charge }+1\\
          \Qt &=\varphit +\theta\eta +\theta^2\Ft &\text{charge }-1
        \end{align}
      \end{subequations}
      They are accompanied by scalar fields $\varphi$ and $\varphit$, the selectrons (or
      selectron and spositron) and by two auxiliary fields $F$ and $\Ft$.

      The photon $A_m$ is placed in a real superfield $V$ together with its superpartner, the
      photino and the auxiliary field $D$. In the Wess--Zumino gauge $V$ reads
      \begin{align}
        V &= -\theta\sigma^m\thetab A_m +\i\theta^2\thetab\lambdab -\i\thetab^2\theta\lambda
        +\tfrac{1}{2}\theta^2\thetab^2 D.
      \end{align}
      
      The gauge transformations we consider are direct generalisations of normal QED, where the
      fields transform as
      \begin{subequations}
        \begin{align}
          \Psi&\to e^{-\i e\alpha(x)}\Psi\\
          A_m &\to A_m + \partial_m \alpha(x).
        \end{align}
      \end{subequations}
      In SQED, the gauge parameter $\alpha$ becomes a chiral superfield, and the gauge
      transformations become
      \begin{subequations}
        \begin{align}
          Q   &\to e^{-\i \alpha}Q\\
          \Qt &\to e^{\i \alpha}\Qt\\
          V &\to V +\tfrac{1}{2}\i\left(\alpha-\alphab\right),
        \end{align}
      \end{subequations}
      so the real superfield $V$ transforms inhomogeneously. This is precisely the property
      that allows us to take $V$ to be in the Wess--Zumino gauge. This procedure not only
      eliminates the lower components of $V$, but also fixes all the additional gauge
      parameters in the superfield $\alpha$ apart from the imaginary part of the scalar
      component which plays the r\^{o}le of the usual gauge parameter of QED. The photino and
      the auxilary field are gauge invariant. 

      The kinetic term for the chiral multiplet as it stands in Eq. (\ref{simplekaehlerpot}),
      however, is not gauge invariant. It has to be modified to
      \begin{align}
        \L_\text{kin,electron} &= \int\d^4\theta\left(\Qb e^{2V}Q +\ol{\Qt}e^{-2V}\Qt\right).
      \end{align}
      The exponential $\exp\{2V\}$ is is actually quite simple in 
      the Wess--Zumino gauge, since
      $V^3=0$. In zeroth order it just reproduces the kinetic terms of
      Eq. (\ref{simplekaehlerpot}); the higher terms mainly amount to
      changing ordinary derivatives to covariant ones, but introduce one important additional
      term that includes the $D$-component of $V$ and is $D\left(\varphib\varphi-
        \bar{\varphit}\varphit\right)$.
      
      We also need kinetic terms for the gauge superfield and, possibly, a superpotential. The
      kinetic term for $V$ is usually written in terms of the field strength superfield (or
      gaugino superfield) $$W_\alpha=-\tfrac{1}{4}\Db^2 D_\alpha V$$ as
      \begin{align}
        \L_\text{kin,SQED}&=\frac{1}{4\mathfrak{e}^2}\int\d^2\theta\,W^\alpha W_\alpha +
        \text{H.c.},
      \end{align}
      where $\mathfrak{e}^2$ is the complexified coupling constant, 
      \begin{align*}
        \frac{1}{\mathfrak{e}^2}&=\frac{1}{e^2} +\i\frac{\vartheta}{8\pi^2},
      \end{align*}
      with $\vartheta$ the vacuum angle. We are using a complex parameter (just as in
      Eq. (\ref{simplesuperpot})) because chiral quantities depend holomorphically on these
      parameters, which can be a powerful tool in many calculations.

      In the Abelian case we deal with now $W_\alpha$ is a gauge invariant chiral superfield.
      Its component expansion is 
      \begin{align}
        W_\alpha&= -\i\lambda_\alpha +\theta^\beta\left(\i
          F_{\alpha\beta}+\varepsilon_{\alpha\beta}D\right)
        +\theta^2\sigma^m_{\alpha\alphad}\partial_m \lambdab^{\alphad},
      \end{align}
      where $F_{\alpha\beta}$ is the photon field strength with spinor indices,
      \begin{align*}
        F_{\alpha\beta}&=\left(\partial_m A_n -\partial_n A_m\right)\sigma^{mn}_{\alpha\beta}
        =F_{\beta\alpha}.
      \end{align*}
      So $\L_\text{kin,SQED}$ takes the component form
      \begin{align}
        \L_\text{kin,SQED}&= \frac{1}{4 e^2}\left[-4\i\lambda\dslash\lambdab +2D^2 -
            F_{mn}F^{mn}\right] +\frac{\vartheta}{64\pi^2}\varepsilon^{mnkl}F_{mn}F_{kl}.
      \end{align}
      
      The superpotential is strongly restricted by gauge invariance. Since we have two
      chiral superfields of opposite charge, the only possible term (restricting to
      renormalisable couplings) is 
      \begin{align}
        \L_\text{SP}&=\int\d^2\theta\,m\Qt Q + \text{H.c.},
      \end{align}
      since the superpotential can only depend on the neutral product $\Qt Q$, and a term
      $\left(\Qt Q\right)^2$ would already require a coefficient with mass dimension -1. Again,
      $m$ is a complex parameter whose absolute value will be the mass of the electron and
      selectron. 

      Now we have kinetic terms for matter and gauge fields as as well as a superpotential.
      What else can be present? Actually, there is one more term, which is possible only for
      Abelian theories, the so-called Fayet-Iliopoulos (or $\xi$) term,
      \begin{align}
        \L_\text{FI}&=2\int\d^4\theta\,\xi V = \xi D
      \end{align}
      which is (super)gauge invariant by itself (this is clearly seen fron Eq. (16c); the
      $\d^4\theta$ integral of chiral and antichiral superfields vanishes). It looks rather
      innocent, but has important consequences as we will see shortly. The coefficient of the
      real superfield, $\xi$,  is real  too. The the dependence on $\xi$ is not holomorphic. 

      Now we can collect all the terms of the Lagrangean,
      \begin{align}
        \begin{split}
          \L&=\L_\text{kin,electron}+\L_\text{kin,SQED}+\L_\text{SP}+\L_\text{FI}\\
          &=\int\d^4\theta\left(\Qb e^{2V}Q +\ol{\Qt}e^{-2V}\Qt+2\xi V\right)\\
          &\quad + \left\{\int\d^2\theta\left(m\Qt Q+\frac{1}{4\mathfrak{e}^2}W^\alpha
              W_\alpha\right) +\text{H.c.}\right\}\\
          &= -D_m\bar{\varphit}D^m \varphit -D_m\varphib D^m \varphi-m\chi\eta -\mb\chib\etab\\
          &\quad -\frac{\i}{2}\left(D_m\chib\sigmab^m\chi -\chib\sigmab^m D_m\chi\right)
          -\frac{\i}{2}\left(D_m\etab\,\sigmab^m\eta -\etab\,\sigmab^m
            D_m\eta\right)\\ 
          &\quad -\frac{1}{e^2}\left(2\i\lambda\dslash\lambdab +F_{mn}F^{mn}\right)
          +\frac{\vartheta}{8\pi}\varepsilon^{mnkl}F_{mn}F_{kl} \\
          &\quad +\Fb F +\ol{\Ft}\Ft +m\varphi\Ft +\mb\varphib\ol{\Ft} +m\varphit F
          +\mb\bar{\varphit}\Fb\\
          &\quad +\frac{2}{e^2}D^2 +\xi D +
          D\left(\varphib\varphi-\bar{\varphit}\varphit\right)\,.
        \end{split}
      \end{align}
      Here, $D_m \alpha=\partial_m\alpha\pm \i A_m\alpha$, where $\alpha=(\varphi,\varphit,
      \chi, \eta)$ and their conjugates and the sign depends on the charge.

      The first two lines contain the kinetic terms for electrons and selectrons and mass terms
      for the fermions, the third line contains the gauge sector kinetic terms and the last two
      lines are the auxiliary field contribution. To obtain mass terms for the scalars and
      exhibit other interesting effects, we have to eliminate these auxiliary fields via their
      equations of motion.  

      The equations of motion are
      \begin{align}
        0&=F  + \mb\bar{\varphit}\, , \\
        0&=\Ft +\mb\varphib \, , \\
        0&=D +\frac{e^2}{4}\left(\xi +\varphib\varphi-\bar{\varphit}\varphit\right)\, ,
      \end{align}
      so the auxiliary field Lagrangean becomes the potential term for the scalar fields, 
      \begin{align}
        \L_\text{aux}&=-\underbrace{\left(|m|^2\bar{\varphit}\varphit
            +|m|^2\varphib\varphi\right)}_{V_F(\varphi,\varphit)}
        -\underbrace{\frac{e^2}{8}\left(\xi
            +\varphib\varphi-\bar{\varphit}\varphit\right)^2}_{V_D(\varphi,\varphit)}
        =-V(\varphi,\varphit).  
      \end{align}
      This is called the scalar potential; it consists of two parts: the $F$ term part and the
      $D$ term part. 

      The scalar potential allows us to analyse the vacua of the theory. We see that the
      potential is a sum of squares, so it can never  be negative. That means that field
      configurations for which $V=0$ are vacua (and supersymmetry is unbroken in these
      cases). However, the existence of supersymmetric vacua depends on the parameters $\xi$
      and $m$. Let us consider four cases:
     
      \paragraph{Case \boldmath$m=\xi=0$:} The scalar potential reduces to
      \begin{align}
        V&=V_D=\left(\varphib\varphi-\bar{\varphit}\varphit\right)^2,
      \end{align}
      which vanishes for any field configuration where $\varphi=\varphit$.
      Both fields, being equal, can take arbitrary  complex value. Thus, the vacuum manifold,
      i.e.\ the set of minima of the scalar potential $V$, is a complex line. What is more
      important is that  distinct  vacua on this line are not physically equivalent. This is
      different from the degenerate vacua situation  in ordinary QFT. Typically the degenerate
      vacua situation takes place when a global symmetry is spontaneously broken. In this case
      one deals with  a compact set of physically equivalent vacua (e.g. the Mexican hat
      potential). In supersymmety non-compact continuous vacuum manifolds are typical. They are
      usually called flat directions. The fields corresponding to these directions are referred
      to as moduli fields. 

      Actually, the requirement $\varphi=\varphit$ is not necessary, since only the absolute
      value enters the potential, and it might seem that there is a larger vacuum
      manifold. However, the relative phase of $\varphi$ and $\varphit$ is a gauge artefact,
      and one can use the product $\varphit\varphi$ as a gauge-independent parametrisation of
      the vacua (sometimes called a composite modulus). 

      All vacua in this case have vanishing potential, so supersymmetry is unbroken, but for
      $\langle\varphi\rangle\neq 0$ or $\langle\varphit\rangle\neq 0$, 
      the gauge symmetry is spontaneously broken. It is clear
      that the vacuum  $\langle\varphi\rangle = 0$ {\em and}
      $\langle\varphit\rangle = 0$, with unbroken gauge symmetry,
      is special.

      \paragraph{Case \boldmath$m=0,\,\xi\neq0$:} In this case $V_F=0$ still holds, but $V_D$
      gets modified, 
      \begin{align}
        V&=V_D=\left(\varphib\varphi-\bar{\varphit}\varphit+\xi\right)^2.
      \end{align}
      If we minimise the potential (assuming\,\footnote{If $\xi<0$, just exchange $\varphi$ and
        $\varphit$ in the following discussion.}  $\xi>0$), we arrive at the condition
      $$\bar{\varphit}\varphit  -\varphib\varphi =\xi \,.$$ 
      This equation has solutions which preserve supersymmetry
      (i.e. $V=0$), but the gauge symmetry is always broken.
      Even if we put
      $\varphi=0$, still $|\varphit|^2=\xi$.  Since in this case the absolute values of the
      fields are fixed, all that is left of the vacuum manifold is a compact $\sf U(1)$-circle,
      with all vacua on this circle being equivalent. Since $\varphi $ need not vanish and its
      absolute value can be arbitrary, the full vacuum manifold is non-compact.

      \paragraph{Case \boldmath$m\neq0,\,\xi=0$:} This case is quite simple, since the
      potential has just one minimum at $\varphi=\varphit=0$, so neither gauge nor
      supersymmetry is broken.   

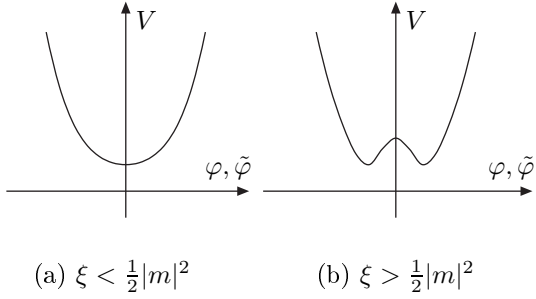
\begin{figure}
        \begin{minipage}{210pt}
          \begin{center}
           
          \subfigure[$\xi<\tfrac{1}{2}|m|^2$]{\label{fayetiliopoulosa}
            \begin{picture}(90,80)(0,0)
              \LongArrow(5,10)(95,10)
              \Text(89,19)[]{\footnotesize $\varphi,\varphit$}
              \LongArrow(50,0)(50,80)
              \Text(58,75)[]{\footnotesize $V$}
              \Curve{(20,70)(25,49.4)(30,36.0)(35,27.8)(40,23.1)(45,20.7)%
                (50,20)(55,20.7)(60,23.1)(65,27.8)(70,36.0)(75,49.4)(80,70)}
            \end{picture}
            }
          \hfill
          \subfigure[$\xi>\tfrac{1}{2}|m|^2$]{\label{fayetiliopoulosb}
            \begin{picture}(100,80)(0,0)
              \LongArrow(0,10)(100,10)\Text(94,19)[]{\footnotesize $\varphi,\varphit$}
              \LongArrow(50,0)(50,80)\Text(58,75)[]{\footnotesize $V$}
              \Curve{(20,70)(25,50)(30,35)(35,24.0)(40,20)(45,26)%
                (50,30)(55,26.0)(60,20.0)(65,24.0)(70,35)(75,50)(80,70)}
            \end{picture}
            }
        \end{center}
        \end{minipage}\hfill
        \begin{minipage}{195pt}
          \caption{Sketch of the scalar potential for SQED with Fayet-Iliopoulos term. If
            \mbox{$\xi<\tfrac{1}{2}|m|^2$}, the minimum of the potential is at the origin, so
            gauge symmetry is unbroken (left figure, case (\ref{fia})). If
            $\xi>\tfrac{1}{2}|m|^2$, the minima are at a nonzero value of some of the fields,
            breaking gauge symmetry (right figure, cases (\ref{fib}) and (\ref{fic})). Since
            $V>0$, supersymmetry is always broken for nonzero $\xi$ and $m$.}
        \end{minipage}
      \end{figure}

      \paragraph{Case \boldmath$m\neq0,\,\xi\neq0$:} 
      In the full potential, it is obvious that
      there will be no supersymmetric vacua, since $V_F$ vanishes only at the origin, where
      $V_D$ does not, and hence $V>0$ always. So supersymmetry is broken. Now we proceed to
      find the minima:
      \begin{subequations}
        \begin{align}
          0&=\frac{\partial V}{\partial \varphib} = 2\left[\left(\varphib\varphi
              -\bar{\varphit}\varphit +\xi\right) +\tfrac{1}{2}|m|^2\right] \varphi\\
          0&=\frac{\partial V}{\partial \bar{\varphit}} = 2\left[-\left(\varphib\varphi
              -\bar{\varphit}\varphit +\xi\right) +\tfrac{1}{2}|m|^2\right] \varphit
        \end{align}
      \end{subequations}
      We distinguish four possibilities:
      \begin{enumerate}[(i)]
        \item \label{fia}$\varphi=\varphit=0$: This is an extremal point of the potential where
          $V=\xi^2$. Whether is is a maximum or minimum depends on the magnitudes of $m$ and
          $\xi$. This can be checked directly, but we will see that when we consider the other
          cases. The potential for the case where this is the minumum is sketched in Figure
          \ref{fayetiliopoulosa}.
        \item \label{fib}$\varphi=0$, $\varphit\neq0$: This means that
          $\tfrac{1}{2}|m|^2-\xi+\bar{\varphit}\varphit=0$, which is only possible if
          $\xi>\tfrac{1}{2}|m|^2$. The value of the potential is
          $V=|m|^2\left(\xi-\tfrac{1}{4}|m|^2\right)$, which is smaller than the value at the
          origin, $V(0)=\xi^2$, if the above condition holds. 
        \item \label{fic}$\varphi\neq0$, $\varphit=0$: Now we need
          $\tfrac{1}{2}|m|^2+\xi+\bar{\varphit}\varphit=0$, which is only possible if
          $\xi<-\tfrac{1}{2}|m|^2$. The potential is in this case
          $V=|m|^2\left(-\xi-\tfrac{1}{4}|m|^2\right)<\xi^2$.  So in this case as in the one
          before, the extremum at the origin is actually not a minimum. The mimima occur at
          finite values of the fields, so gauge symmetry is broken in the vacuum. The potential
          in these cases is sketched in Figure \ref{fayetiliopoulosb}.
        \item $\varphi\neq 0$, $\varphit\neq 0$: In this case, there is no extremal point at
          all for nonzero mass. If $m=0$, it reduces to the case considered before where gauge
          symmetry is broken, but supersymmetry is not.
      \end{enumerate}

%%%%%%%%%%%%%%%%%%%%%%%%%%%%%%%%%%%%%%%%%%%%%%%%%%%%%%%%%%%%%%%%%%%%%%%%%%%%%%%%%%%%%%%%%%
%%%%%%%%%%%%%%%%%%%%%%%%%%%%%%%%% Figure of the scalar potential with FI term %%%%%%%%%      

%%%%%%%%%%%%%%%%%%%%%%%%%%%%%%%%%%%%%%%%%%%%%%%%%%%%%%%%%%%%%%%%%%%%%%%%%%%%%%%%%%%%%%%%%%
%%%%%%%%%%%%%%%%%%%%%%%%%%%%%%%%%%%%%%%%%%%%%%%%%%%%%%%%%%%%%%%%%%%%%%%%%%%%%%%%%%%%%%%%%%

    \subsection{SUSY Gluodynamics}
      After SQED, we will now consider non-Abelian gauge theories. The simplest example is a
      pure Super-Yang-Mills theory (SYM) without matter, sometimes called SUSY
      gluodynamics. This theory contains a vector field $A_m=A_m^a T^a$ and a Majorana spinor
      $\lambda_\alpha=\lambda_\alpha^a T^a$, both in the adjoint representation of the gauge
      group with generators $T^a$. They are again grouped in a vector superfield $V=V^a T^a$
      which we can take to be in 
      the Wess--Zumino gauge. The action looks rather simple,
      \begin{align}
        \begin{split}\label{symaction}
          S&=\frac{1}{2\mathfrak{g}^2}\int\d^4 x \d^2\theta \tr W^\alpha W_\alpha +
          \text{H.c.}\\ 
          &= \int\d^4 x \left\{ -\frac{1}{4g^2}G_{mn}^a G^{a\,mn}
            +\frac{\vartheta}{32\pi^2}{G}_{mn}^a \widetilde{G}^{a,mn}
            +\frac{1}{g^2}\lambdab^a \Dslash \lambda^a\right\},
        \end{split}
      \end{align}
      where $D_m=\partial_m-\i A_m$ is the covariant derivative and $\mathfrak{g}$ is the
      complexified coupling constant,
      \begin{align*}
        \frac{1}{\mathfrak{g}^2}&=\frac{1}{g^2}+\i\frac{\vartheta}{8\pi^2}.
      \end{align*}
      Moreover,
      $$
        \widetilde{G}^{a,mn} =\frac{1}{2}\,  \epsilon^{mnkl} \, G^a_{kl}\,.
      $$
      This theory is believed to be confining and is asymptotically free, i.e.\ the first
      coefficient of the $\beta$-function (Gell-Mann--Low function) is negative. For a
      $\mathsf{SU(}N\mathsf)$ gauge group, $\beta_0=-3N$.\footnote{This can be understood from
        the $\beta$ function of a non-supersymmetric $\mathsf{SU(}N_\text{\sc c}\mathsf)$ gauge
        theory with $N_\text{\sc f}$ flavours, where $\beta_0=-\frac{11}{3}N_\text{\sc c}
        +\frac{2}{3}N_\text{\sc f}$. The supersymmetric theory can be thought of as a usual
        theory with one gaugino, so $N_\text{\sc f}=1$. There is an additional factor of
        $N_\text{\sc c}$ for the gaugino contribution because it is in the adjoint
        representation of the gauge group rather than in the fundamental. The graphs
        contributing to this term contain a Casimir operator for the representation, and the
        ratio of the Casimirs in the adjoint and fundamental representation is $2N_\text{\sc
          c}$. Furthermore, the gaugino is a Majorana fermion with only half as many degrees of
        freedom as the Dirac one, so there is another factor of $\frac{1}{2}$, resulting in
        $\beta_0=-\frac{11}{3}N_\text{\sc c} +\frac{2}{3}N_\text{\sc c}=-3N_\text{\sc c}$.}

      \subsubsection{Gluino Condensate}
        The aspect we will focus on is the gluino condensate
        $\langle\lambda^a\lambda^a\rangle$, which vanishes perturbatively, but does not vanish
        in the nonperturbative treatment  and can be computed {\em exactly}
        \cite{shifmanvainshtein}. We will calculate its value in Chapter \ref{calcchap}. Now we
        give the result and argue for its plausibility. The result is 
        \begin{align}\label{gluinocondensate}
          \left\langle\lambda^{a,\alpha}\lambda^a_\alpha\right\rangle&=-6N \Lambda^3
          \exp\left\{\i\left(\frac{2\pi k}{N}+ \frac{\vartheta}{N}\right)\right\}\,,
          \qquad k=0,1, ..., N-1.
        \end{align}
        In this expression, $\Lambda$ is the scale where the theory becomes strongly coupled. It
        is the only dimensionful parameter, so it has to appear to the third power. The
        pre-factor $N$ can be understood since the sum over group indices involves $N^2$ terms,
        but a factor of $g$ is included in $\lambda$, and $g^2\sim N^{-1}$. In the exponential,
        the parameter $k$, $k=0,\dotsc,N-1$ labels distinct supersymmetric vacua, reflecting
        the $N$-fold vacuum degeneracy.

        The above general form can be derived from holomorphicity in conjunction with
        renormalization  group (RG) arguments. The only RG invariant expression of dimension of
        mass which can be built from $g$ and the renomalisation scale $\mu$ is \linebreak
        $\mu\exp\{-8\pi^2(|\beta_0|g)^{-1}\}$, so  
        \begin{align}
          \begin{split}
            \left\langle\lambda\lambda\right\rangle&=c
            \left[\mu \exp\left\{-\frac{8\pi^2}{|\beta_0|}\frac{1}{g^2}
            \right\}\right]^3 \\
            &=c \,\mu^3 \exp\left\{-\frac{8\pi^2}{N}\frac{1}{g^2}\right\}
          \end{split}
        \end{align}
        where $c$ is a constant and the second equality holds for $\mathsf{SU(}N\mathsf)$ gauge
        groups \linebreak(\mbox{$\beta_0=-3N$}). On the other hand, $\langle\lambda\lambda\rangle$ is a
        chiral quantity, so it must depend analytically on $\mathfrak{g}$, not just on $g$, see
        \cite{shifmanvainshtein}, 
        \begin{align}
          \begin{split}\label{condensate1}
            \left\langle\lambda\lambda\right\rangle&=c' \mu^3
            \exp\left\{-\frac{8\pi^2}{N}\left(\frac{1}{g^2}+\i\frac{\vartheta}{8\pi^2}\right)
            \right\}=c'\mu^3\exp\left\{-\frac{8\pi^2}{N}\frac{1}{g^2}\right\}
            \exp\left\{\i\frac{\vartheta}{N}\right\}
          \end{split}
        \end{align}
        with a possibly different constant $c'$. Next we notice that physical results must be
        $2\pi$-periodic in the vacuum angle $\vartheta$.  The above result  is not, so we have
        to change the last factor in Eq. (\ref{condensate1}) to 
        \begin{align*}
          \exp\left\{\i\left(\frac{2\pi k}{N}+ \frac{\vartheta}{N}\right)\right\}.
        \end{align*}
        The parameter $k$ labels the distinct (but degenerate) vacua, of which there are $N$
        This is exactly the value of  the Witten index $I_W$ in $\mathsf{SU(}N\mathsf)$
        theories \cite{wittenindex}, which in fact counts the number of vacua. If there are no
        matter fields, supersymmetry is always unbroken.

      \subsubsection{Existence of {\boldmath$N$} Vacua}
        We can also deduce the number of vacua by considering a global symmetry of the
        Lagrangean (\ref{symaction}) (besides   supersymmetry). There is a global chiral
        $\mathsf{U(1)}$ symmetry which acts as
        \begin{subequations}
        \begin{align}
          \lambda  &\to e^{\i\alpha}\lambda \,,\\
          \lambdab &\to e^{-\i\alpha} \lambdab\,.
        \end{align}
        \end{subequations}
        This global symmetry corresponds to the current $J^m=\lambdab\sigmab^m\lambda$ which is
        classically conserved, $\partial_m J^m=0$.

%%%%%%%%%%%%%%%%%%%%%%%%%%%%%%%%%%%%%%%%%%%%%%%%%%%%%%%%%%%%%%%%%%%%%%%%%%%%%%%%%%%%%%%%%%%%%
%%%%%%%%%%%%%%%%%%% Triangle Anomaly figure %%%%%%%%%%%%%%%%%%%%%%%%%%%%%%%%%%%%%%%%%%%%%%%%%
     
        \begin{wrapfigure}[10]{r}{80pt}
          \begin{center}
          \begin{picture}(80,90)(0,0)
            \SetWidth{1} %bold X for the J_m coupling
              \Line(37,103)(43,97)\Line(37,97)(43,103)
            \SetWidth{.5}
            \Line(40,100)(40,70)
            \Text(42,85)[l]{$J_m$}%% J_m line
            \Vertex(40,70){2}
            \ArrowLine(40,70)(20,35.4)
            \Vertex(20,35.4){2}
            \ArrowLine(20,35.4)(60,35.4)
            \Vertex(60,35.4){2}
            \ArrowLine(60,35.4)(40,70)
            \Gluon(20,35.4)(20,5){3}{4}
            \Text(15,23)[r]{$\widetilde{G}$}          
            \Gluon(60,35.4)(60,5){3}{4}
            \Text(65,20)[l]{$G$}
          \end{picture}
          \caption{Triangle anomaly graph.\label{triangleanomaly}}
          \end{center}
        \end{wrapfigure}
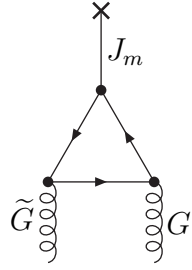

%%%%%%%%%%%%%%%%%%%%%%%%%% End figure %%%%%%%%%%%%%%%%%%%%%%%%%%%%%%%%%%%%%%%%%%%%%%%%%%%%%%%
%%%%%%%%%%%%%%%%%%%%%%%%%%%%%%%%%%%%%%%%%%%%%%%%%%%%%%%%%%%%%%%%%%%%%%%%%%%%%%%%%%%%%%%%%%%%%%

        However, in fact,  this current is not conserved because of an anomaly originating from
        the triangle diagram in Figure (\ref{triangleanomaly}). This means that the current is
        no longer conserved on the quantum level. Its divergence can be calculated exactly, 
        \begin{align}\label{e:triangleanomaly}
          \partial_m J^m&=\frac{N}{16\pi^2} G^a_{mn}\widetilde{G}^{a,mn}
        \end{align}
        Thus, the chiral $\mathsf{U(1)}$ is not a symmetry of the theory. The anomaly
        explicitly breaks the chiral $\mathsf{U(1)}$ down to a discrete subgroup, namely the
        $\mathbb{Z}_{2N}$  corresponding to $\alpha=\pi k/N$,  \mbox{$k=1,\dotsc,2N$}. 
 
        This is not the end of the story. A nonvanishing $\langle\lambda\lambda\rangle$ is not
        invariant even under this $\mathbb{Z}_{2N}\simeq \mathbb{Z}_N\times\mathbb{Z}_2$, but
        breaks it further down to $\mathbb{Z}_2$. This spontaneous breaking of the discrete
        symmetry implies that we are left with $N$ distinct (discrete) vacua.

    \subsection{Calculation of the Gluino Condensate\label{calcchap}}
    
      Now we will actually calculate the gluino condensate. To do this head on in the strong
      coupling regime is, however, impossible. To reach the result, we will have to take a
      rather complicated detour to stay with theories that are calculable, like like in
      crossing an ocean by jumping from small island to small island. One false step and we
      will be lost in the sea of noncalculability.
      We will go through the following steps:
      \begin{itemize}\setlength{\itemsep}{0cm}
        \item Restrict to $\sf SU({\rm 2})$ gauge group: This is mainly for technical
          simplicity, and since $\langle\lambda\lambda\rangle\sim N$, the result generalises to
          any ${\sf SU(}N\sf)$.
        \item Introduce one flavour of matter fields (fermions and sfermions): They will
          generically break the gauge symmetry, making the gauge fields massive. The remaining
          (massless) fields can ge assembled in one chiral superfield. 
        \item Go to the low energy effective theory: Since we are interested in the gluino
          condensate and hence in the vacuum structure, we can integrate out the heavy
          modes. We will find that instantons generate a nonperturbative superpotential. The
          resulting scalar potential is a runaway potential, i.e. the minima are at infinite
          field values. We will cure this by adding a small mass term for the formerly massless
          fields. 
        \item At the next step, we will relate the vacuum expectation value of the light scalar
          fields to the gluino condensate via the Konishi anomaly. The result will still depend
          on the mass introduced before.
        \item Finally, we take the mass to infinity (to get rid of the matter fields we
          introduced in the beginning). Luckily, we have full control over the extrapolation
          procedure, the gluino condensate stays finite in this  limit, and eventually we
          arrive at the result given in Eq. (\ref{gluinocondensate}). 
      \end{itemize}
      
      \subsubsection{One-Flavour SQCD}
        From now on, we will consider only $\sf SU(2)$ as gauge group. The generators are given
        by the Pauli matrices, $T^a=\frac{1}{2}\tau^a$, $a=1,2,3$. We also add one flavour
        of matter, i.e. a Dirac spinor in the fundamental representation and its scalar
        superpartners. The fields can be assembled in two chiral superfields $Q^i$, $\Qt^j$
        transforming as doublets under $\sf SU(2)$.\footnote{Actually, they should transform as
          doublet and antidoublet, but for $\sf SU(2)$ the antidoublet is isomorphic to the
          doublet (via $\Qt^i=\varepsilon^{ij}\Qt_j$).} For notational brevity, we will
        write them as $Q^i_f$ with a ``subflavour'' index $f=1,2$ to denote both $Q^i\equiv
        Q^i_1$ and $\Qt^i\equiv Q^i_2$. The component fields are called $\phi_f$ (scalars),
        $\psi_f$ (fermions) and $F_f$ (auxiliaries). The matter Lagrangean is just the kinetic
        term, we will add a mass term only later,
        \begin{align}
          \begin{split}
            \L_\text{matter}&=\int\d^4\theta\left\{\Qb e^{2V} Q +\ol{\Qt} e^{2V}\Qt\right\}=
            \int\d^4\theta\,\Qb_f e^{2V} Q_f\\
            &= -\psib_f\i\Dslash\psi_f -D^m\phib_f D_m \phi_f
            +\tfrac{\i}{\2}\left[\phib_f\left(\lambda\psi_f\right)-\text{H.c.}\right]\\ 
            &\quad +D^a \phib_f T^a \phi_f +\Fb_f F_f .\\
          \end{split}
        \end{align}
        Since there is no superpotential so far, we can forget about the $F$-terms. The
        $D$-term, however, gives a contribution to the scalar potential after elimination of
        $D^a$ through its equation of motion (there is a $D^2$ term from the kinetic
        Lagrangean of the gauge field), 
        \begin{align}
          \begin{split}
            V&=V_D= \frac{1}{g^2}\left(\phib_f T^a \phi_f\right)^2.
          \end{split}
        \end{align}
        This potential vanishes for the following field values:
        \begin{align}
          \phi_1&= v\left(\begin{array}{c} 1\\0\end{array}\right)\,,  & \phi_2&=
          v\left(\begin{array}{c} 0\\1\end{array}\right),
        \end{align}
        where $v$ is a complex number. {\em A priori}, $\phi_1$ and $\phi_2$ could have
        different phases, but the relative phase can be gauged away. That the potential indeed
        vanishes can be seen from the explicit form of the generators:  
        \begin{itemize}
          \item $a=1,2$: The generators are off-diagonal, and the product of the form
            \begin{align*}
              \left(1,0\right)\left(\begin{array}{cc} 0 &
                  X\\Y&0\end{array}\right)\left(\begin{array}{c} 1\\0\end{array}\right)  
            \end{align*}
            vanishes for any $X$, $Y$.
          \item $a=3$: $\tau^3$ is diagonal, but traceless, and hence
            \begin{align*}
              \phi_1\tau^3\phi_1 + \phi_2 \tau^3 \phi_2&= v^2 -v^2=0.
            \end{align*}
        \end{itemize}
        
        So there is a non-compact vacuum manifold parameterised by $v$. The definition of $v$
        in Eq. (39) seems to be gauge dependent. In fact, the degree of gauge dependence is
        minimal, since $v^2$ is gauge independent. This is clearly seen from 
        \begin{align}
          v^2&=\tfrac{1}{2}\phi^i_f \phi^j_g \varepsilon_{ij} \varepsilon^{fg}\equiv H^2.
        \end{align}
        Hence the flat direction is a one-dimensional complex line. For any value of $v^2\neq
        0$, the gauge symmetry is broken and the gauge bosons acquire the mass $M_g=g v$,
        proportional to the as yet undetermined parameter $v$. 

        We will later see that the theory actually wants to develop a large $v$. If then
        $M_g\gg 1$, the theory is weakly coupled and we can actually calculate everything. The
        gluons thus behave rather like $W$ bosons.  

      \subsubsection{Low-Energy Theory and Nonperturbative Superpotential}
        The gauge symmetry breaking and Higgs mechanism reshuffle the degrees of
        freedom, since the vector bosons eat some part of the scalar fields 
        acquiring longitudinal components. The degrees of freedom before and after gauge
        symmetry breaking are listed in Table \ref{dof-table}. 
        
%%%%%%%%%%%%%%%%%%%%%%%%%%%%%%%%%%%%%%%%%%%%%%%%%%%%%%%%%%%%%%%%%%%%%%%%%%%%%%%%%%%%%%%%%%%%%%%%%%%%
%%%%%%%%%%%%%%%%%%%%%%%%%%% Table Symmetry breaking %%%%%%%%%%%%%%%%%%%%%%%%%%%%%%%%%%%%%%%%%%%%%%%%
        \begin{table}
          \begin{center}
            \begin{tabular}{||c|l|c|c||c|l|c|c|c||}\hhline{|t:*4=:t:*5=:t|}
              \multicolumn{4}{||c||}{before \rule[-7pt]{0pt}{23pt}} &
              \multicolumn{5}{c||}{after}\\\hhline{||*4-||*{5}-||}
              SF & Component field & b& f &SF &Component field & b & f &$M$\\\hhline{||*4-||*{5}-||}
              \multirow{2}{*}{$V$}& 3 gluons & 6&  &  \multirow{4}{*}{$V_m$} &3 $W$ bosons & 9
              &&\hspace{-3mm}\ldelim\}{4}{4mm}\multirow{4}{*}{$M_g$}\\
              & 3 gluinos & & 6&& 3 gluinos &&6&\\\hhline{||*{4}-||}
              \multirow{2}{*}{$Q$} & 2 complex scalars & 4 &&& 3 gauge fermions &&6&\\
              & 2 Weyl fermions &&4&&3 real scalars &3&&\\\hhline{||*4-||*{5}-||}
              \multirow{2}{*}{$\Qt$} & 2 complex scalars &4 &&\multirow{2}{*}{$X$} & 1 complex
              scalar &2&&\multirow{2}{*}{0}\\
              & 2 Weyl fermions &&4&& 1 Weyl fermion &&2&\\\hhline{|b:*{4}=:b:*5=:b|}
            \end{tabular}
            \caption{\label{dof-table} Degrees of freedom before and after gauge symmetry
              breaking. In the SF column, we give the superfield and in the next column the
              contained component fields. b and f denote bosonic and fermionic degrees of
              freedom, respectively, and $M$ is the mass of the fields. Before symmetry
              breaking, all fields are massless. $V_m$ denotes a massive vector superfield,
              including the additional fermion and scalar, and $X$ is a chiral multiplet which
              includes the fields which are still massless after the Higgs mechanism.}
          \end{center}
        \end{table}
%%%%%%%%%%%%%%%%%%%%%%%%%%%%%%%%%%%%%%%%%%%%%%%%%%%%%%%%%%%%%%%%%%%%%%%%%%%%%%%%%%%%%%%%%%%%%%%%%%%%%%
%%%%%%%%%%%%%%%%%%%%%%%%%%%%%%%%%%%%%%%%%%%%%%%%%%%%%%%%%%%%%%%%%%%%%%%%%%%%%%%%%%%%%%%%%%%%%%%%%%%%%%

        To analyse the vacuum structure, we can integrate out the heavy fields and consider just
        the low-energy effective action. All we are left with is a chiral multiplet $X$ defined
        by 
        the formula
        $$X^2=Q^i_f Q^j_g \varepsilon_{ij} \varepsilon^{fg}\,.$$ Its scalar component is the
        Higgs field $H\equiv v$, and there  are   superpartners,
        \begin{align}
          X&=H +\2\theta\psi +\theta^2 F_X\,.
        \end{align}

        For the resulting Lagrangean, we get the usual kinetic term and nothing else, at  the
        perturbative level. However, instantons induce a nonperturbative contribution to the
        superpotential. This contribution is severely restriced by the non-anomalous
        $R$-symmetry of the theory, on which I will dwell momentarily. The only allowed form is
        \begin{align}\label{nonperturbativesuperpotential}
          \L&=\int\d^4\theta\,\ol{X}X + \left\{c \int\d^2\theta\frac{\Lambda^5}{X^2}
            +\text{H.c.}\right\}.
        \end{align}
        The nonperturbative nature of the superpotential can be seen from its dependence on
        $\Lambda^5$, as opposed to perturbative terms which could contain only powers of  $
        (\ln\Lambda )^{-1}$. Division by $X$ is to be understood as an expansion in $\theta$, 
        \begin{align}
          \begin{split}\label{oneoverxsquaredexpansion}
            \frac{1}{X^2}&=\frac{1}{H^2+2\2\theta\chi H +\theta^2\left(2FH-\chi\chi\right)}\\
            &=\frac{1}{H^2}-\frac{1}{H^3}\left(2\2\theta\chi +2\theta^2F\right)
            -\frac{1}{H^4}\theta^2\chi\chi.
          \end{split}
        \end{align}

        Now I return to the issue of the $R$-symmetry, which is a global (chiral) $\sf U(1)$
        symmetry of the fundamental theory with respect to rotations of  the fermions and
        sfermions. $R$-symmetry does not commute with supersymmetry, as it requires a rotation
        of the Grassmann parameters $\theta$ and $\bar\theta$. Assume that $\d\theta \to
        e^{\i\alpha}\d\theta$. (Naturally $\d\bar\theta \to e^{-\i\alpha}\d\bar\theta$.) Since
        the kinetic term of the gauge fields is proportional to $\int {\rm tr}\, W^2
        \d^2\theta$ the invariance of the Lagrangean then requires that $W\to e^{-\i\alpha}W$
        which entails, in turn that $\lambda \to e^{-\i\alpha}\lambda $. 

        The $\sf U(1)$  charges of the matter fermions are determined by anomaly
        cancellation: we want the $R$ current to be conserved not only classically, but,
        rather, exactly conserved. The exactly conserved $R$ current is
        $$\lambdab\sigmab^m\lambda - 2\sum_f \, \bar\psi^f \sigmab^m\psi_f\,.$$ 
        The emergence of the factor 2 in front of the matter fermions is explained by the fact
        that for $\sf SU(2)$ the contributions to the anomaly of the gaugino is twice  larger
        that that of the matter fermions, see footnote before Eq. (32). Therefore, from the law
        of the $R$ rotation of $\lambda$ above we deduce that $\psi_f \to
        e^{2\i\alpha}\psi_f$. What remains to be established is the $R$ charge of the sfermion
        field $\phi_f$. The easiest way is to examine the vertex $(\psi_f\lambda )\bar\phi^f$
        in the Lagrangean. Its invariance requires that $\bar\phi^f\to e^{-\i\alpha}
        \bar\phi^f$, which implies that $ \phi^f\to e^{ \i\alpha}  \phi^f$. Finally, let us
        note that if  $\d\theta \to e^{\i\alpha}\d\theta$ then $ \theta \to
        e^{-\i\alpha}\theta$. Combining this with the above results we conclude that the matter
        superfields are transformed as $Q^i_f\to e^{\i\alpha} Q^i_f$ while  $V\to V$. The
        $R$-charges of the fields are collected in table \ref{r-table}. The $R$-charge $n$ means
        that a field $\phi$ transforms as $\phi\to e^{\i n\alpha}\phi$. 
         
        \addtolength{\columnsep}{5pt}
        \begin{wraptable}[5]{r}{149pt}
          \vspace{-7pt}
          \begin{tabular}{c|ccccc}
            $\phi$ & $\lambda$ & $\varphi_f$ & $\psi_f$ & $\theta$ & $\d\theta$\\\hline
            $n$ & $-1$ &$1$ &$2$ &  $-1$ & $1$
          \end{tabular}
          \caption{\label{r-table} $R$-charges of the fields.}
        \end{wraptable}
        \addtolength{\columnsep}{-5pt}
                
        The non-anomalous $R$-symmetry must be conserved in the low-energy effective action
        which emerges after integration over the heavy fields. Note that $X\to e^{\i\alpha}X$;
        in other words, the $R$ charge of $X$ is unity. Since the kinetic terms are $\sim
        \ol{X}X$, there is no restriction from $R$-symmetry on the kinetic term. The
        superpotential is quite constrained, however. Indeed, if we impose analyticity ($W\sim
        X^k$), we find  
        \begin{align}
          \int\underbrace{\d^2\theta}_\text{\clap{charge 2}}\,X^k&\to
          e^{\i(2+k)\alpha}\int\d^2\theta\,X^k,
        \end{align}
        so the superpotential must be $\sim X^{-2}$. To guarantee the appropriate  mass
        dimension, we then have to insert the fifth power of $\Lambda$, the only dimensionful
        parameter of the theory. Note that exactly $\Lambda^5$ emerges from the one-instanton
        measure. 
        
        Now, the superpotential is completely fixed up to a coefficient. This coefficient, in
        principle, could be zero. Therefore, strictly speaking, we do not yet know whether a
        superpotential is generated nonperturbatively, but if it is, it must be of the form
        given in Eq. (\ref{nonperturbativesuperpotential}).  
        
      \subsubsection{Instantons Generate this Superpotential}      
        We will now check that instantons do generate this superpotential. Instantons (at weak
        coupling) are classical solutions to the gauge field equations of motion with
        non-trivial topology in field space and finite action $S_*\neq 0$. Their contributions
        are suppressed by $\exp\{- S_*\}= \exp\{- 8\pi^2/g^2 \} $, and so perturbative effects
        (corresponding to $S=0$) are dominant. But in the case of the superpotential, there are
        no perturbative contributions (this is the famous non-renormalisation theorem for
        superpotentials), so the instanton effects decide the outcome. 
        
        Our model is constructed in such a way that at $v\neq 0$ the gauge symmetry of the
        model is completely broken and {\em all} gauge bosons become massive (in fact, very
        heavy if $v$ is large). Why is it that the requirement of the complete breaking of the
        gauge symmetry is so important?  
        
        If all gauge bosons are  heavy (much heavier than $\Lambda$) the theory is at weak
        coupling. This means that all calculations, including instanton calculations, are under
        complete control and are reliable. If, on the other hand, a non-Abelian subgroup
        remains unbroken,the analysis is dragged, with necessity, into a strong coupling
        domain, where theoretical control is lost. Technically this loss manifests itself in
        the fact that, with  unbroken non-Abelian  gauge group, integrals over the instanton
        size become divergent, and there is no way one can make instanton calculation
        well-defined.

%%%%%%%%%%%%%%%%%%%%%%%%%%%%%%%%%%%%%%%%%%%%%%%%%%%%%%%%%%%%%%%%%%%%%%%%%%%%%%%%%%%%%%%%%%
%%%%%%%%%%%%%%%%%%%%%%%%%%%% Instanton diagrams %%%%%%%%%%%%%%%%%%%%%%%%%%%%%%%%%%%%%%%%%%
        \begin{figure}
          \begin{center}\scriptsize
            \subfigure[\sl Generic Instanton with all zero modes]{\label{genericinstanton}
              \begin{picture}(100,100)(0,0)\SetOffset(50,50)
                %% The Instanton
                \BCirc(0,0){15}
                \Text(0,0)[]{$I$}
                %% The Fermion Lines, four lambdas and two psis
                % Right side
                \ArrowLine(14.7721, 2.60472)(39.3923, 6.94593)
                \Text(32,11)[]{$\lambda$}
                \ArrowLine(11.4907, 9.64181)(30.6418, 25.7115)
                \Text(25,30)[]{$\lambda$}
                \ArrowLine(12.9904, -7.5)(34.641, -20.)
                \Text(25,-21)[]{$\psi_1$}
                % Left side
                \ArrowLine(-14.7721, 2.60472)(-39.3923, 6.94593)
                \Text(-32,11)[]{$\lambda$}
                \ArrowLine(-11.4907, 9.64181)(-30.6418, 25.7115)
                \Text(-25,30)[]{$\lambda$}
                \ArrowLine(-12.9904, -7.5)(-34.641, -20.)
                \Text(-25,-21)[]{$\psi_2$}
              \end{picture}
              }\hspace{1cm}
            \subfigure[\sl Four zero modes eliminated]{\label{instantonlambdalambda} 
              \begin{picture}(100,100)(0,0)\SetOffset(50,50)
                %% The Instanton
                \BCirc(0,0){15}
                \Text(0,0)[]{$I$}
                %% The free fermions
                \ArrowLine(14.0954, 5.1303)(37.5877, 13.6808)
                \Text(30,17)[]{$\lambda$}
                \ArrowLine(-14.0954, 5.1303)(-37.5877, 13.6808)
                \Text(-30,17)[]{$\lambda$}
                %% The lower Fermion lines linked to the VEV
                %% Right Side
                \ArrowLine(14.0954, -5.1303)(28.1908, -10.2606) %% lambda
                \Text(33,-6)[]{$\lambda$}
                \ArrowLine(9.64181, -11.4907)(19.2836, -22.9813)%% Psi
                \Text(18,-29)[]{$\psi$}
                \CArc(25.4415, -17.8143)(8.03848,-140,70)
                \Vertex(32.0262, -22.425){1}
                \Line(32.0262, -22.425)(41.8245, -29.2858)%% phi
                \Text(39,-22)[]{$\phi$}
                \Text(41.8245, -29.2858)[]{\boldmath \scriptsize$\times$}
                %% Left Side
                \ArrowLine(-14.0954, -5.1303)(-28.1908, -10.2606)%% lambda
                \Text(-33,-6)[]{$\lambda$}
                \ArrowLine(-9.64181, -11.4907)(-19.2836, -22.9813)%% Psi
                \Text(-18,-29)[]{$\psi$}
                \CArc(-25.4415, -17.8143)(8.03848,110,320)
                \Vertex(-32.0262, -22.425){1}
                \Line(-32.0262, -22.425)(-41.8245, -29.2858)%% phi
                \Text(-39,-22)[]{$\phi$}
                \Text(-41.8245, -29.2858)[]{\boldmath \scriptsize$\times$}
              \end{picture}
              }\hspace{1cm}
            \subfigure[\sl Remaining $\lambda$ zero modes converted to $\psi$'s]{\label{instantonpsipsi}
              \begin{picture}(100,100)(0,0)\SetOffset(50,50)
                %% The Instanton
                \BCirc(0,0){15}
                \Text(0,0)[]{$I$}
                %% The upper fermions linked to the vev
                % Right side
                \ArrowLine(14.0954, 5.1303)(28.1908, 10.2606) %lambda
                \Text(21,13)[]{$\lambda$}
                \Vertex(28.1908, 10.2606){1}
                \ArrowLine(28.1908, 10.2606)(45.1908, 10.2606) %psi
                \Text(43,15)[]{$\psi$}
                \Line(28.1908, 10.2606)(28.1908,30.2606)%phi
                \Text(32,22)[]{$\phi$}
                \Text(28.1908,30.2606)[]{\boldmath \scriptsize$\times$}
                % Left side
                \ArrowLine(-14.0954, 5.1303)(-28.1908, 10.2606) % lambda
                \Text(-21,13)[]{$\lambda$}
                \Vertex(-28.1908, 10.2606){1} 
                \ArrowLine(-28.1908, 10.2606)(-45.1908, 10.2606)%psi
                \Text(-43,15)[]{$\psi$}
                \Line(-28.1908, 10.2606)(-28.1908,30.2606)%phi
                \Text(-32,22)[]{$\phi$}
                \Text(-28.1908,30.2606)[]{\boldmath \scriptsize$\times$} 
                %% The lower Fermion lines linked to each othe vie the VEV
                \ArrowLine(14.0954, -5.1303)(28.1908, -10.2606) %% lambda
                \Text(33,-6)[]{$\lambda$}
                \ArrowLine(9.64181, -11.4907)(19.2836, -22.9813)%% Psi
                \Text(18,-29)[]{$\psi$}
                \CArc(25.4415, -17.8143)(8.03848,-140,70)
                \Vertex(32.0262, -22.425){1}
                \Line(32.0262, -22.425)(41.8245, -29.2858)%% phi
                \Text(39,-22)[]{$\phi$}
                \Text(41.8245, -29.2858)[]{\boldmath \scriptsize$\times$}
                \ArrowLine(-14.0954, -5.1303)(-28.1908, -10.2606)%% lambda
                \Text(-33,-6)[]{$\lambda$}
                \ArrowLine(-9.64181, -11.4907)(-19.2836, -22.9813)%% Psi
                \Text(-18,-29)[]{$\psi$}
                \CArc(-25.4415, -17.8143)(8.03848,110,320)
                \Vertex(-32.0262, -22.425){1}
                \Line(-32.0262, -22.425)(-41.8245, -29.2858)%% phi
                \Text(-39,-22)[]{$\phi$}
                \Text(-41.8245, -29.2858)[]{\boldmath \scriptsize$\times$}
              \end{picture}
              }
            \caption{Instanton contribution to the superpotential. Fermionic zero modes are
              convoluted and converted using the vacuum expaectation value of $\phi$
              \label{f:instantons}} 
          \end{center}
        \end{figure}
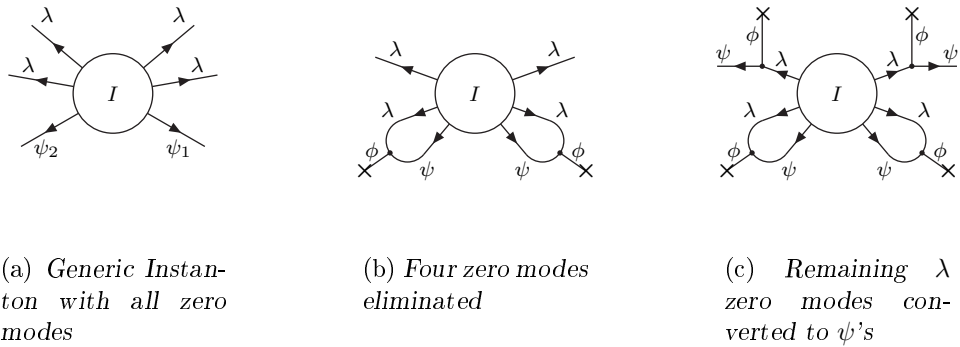
%%%%%%%%%%%%%%%%%%%%%%%%%%%%%%%%%%%%%%%%%%%%%%%%%%%%%%%%%%%%%%%%%%%%%%%%%%%%%%%%%%%%%%%%%%
%%%%%%%%%%%%%%%%%%%%%%%%%%%%%%%%%%%%%%%%%%%%%%%%%%%%%%%%%%%%%%%%%%%%%%%%%%%%%%%%%%%%%%%%%%
         
        After this remark let us pass to  determination of the coefficient $c$ in the model at
        hand. Usually, instanton analyses are quite cumbersome. Supersymmetry makes it  rather
        trivial, however. 

        The instanton suppression factor $\exp\{- 8\pi^2/g^2 \}$ quoted above is just the
        classical instanton action. To carry out the actual calculation we must integrate over
        fluctuations of all fields over the instanton solution. To this end one expands the
        field fluctuations in modes. There are a few zero modes, and an infinite number of
        non-zero modes. Handling non-zero modes is the most labour-intensive and time-consuming
        part of the problem. In supersymmetric theories one can prove, however, that all
        nonzero modes cancel each other, so the instanton analysis reduces to that of zero
        modes \cite{novikov}. 

        Pictorially, the  instanton is represented by a circle with an $I$ in it in
        Figure~\ref{f:instantons}. In  the model under consideration the  instanton has four 
        gaugino zero modes and two zero modes corresponding to two subflavors of the matter
        fermions (see Fig.~a). In order to determine the number of the zero modes one does not
        have to actually find them. The anomaly relation (\ref{e:triangleanomaly}), combined with 
        $$
          \int d^4 x \frac{1}{16\pi^2}\left( G_{mn}^a \tilde{G}^{a,\, mn}\right)_{\rm inst} =
          2,
        $$
        fixes the number of the gaugino zero modes, while a similar anomaly relation for the
        matter fermion axial current determines the number of the matter fermion zero
        modes. All fermion lines are directed outward, so the instanton breaks the chiral
        symmetry. It makes all anomalies explicit. The chiral currents of both gauginos and
        matter fermions are anomalous. As I have already mentioned, however,  the combination
        $j_\lambda -2j_\psi$ is not.  
        
        At first sight, it seems that the fact that there are six fermion zero modes in the
        instanton background forbids generation of the effective Lagrangean (42) (see also
        Eq. (43)). Indeed, this effective Lagrangean requires two and only two fermion zero
        modes, since the only term surviving after integration over $\d^2\theta$ is the last
        one in Eq. (43). Let us not rush to hasty  conclusions, however.

        The fundamental Lagrangean of our model contains the vertex $(\psi_f\lambda
        )\bar\phi^f$. It allows one to convolute a pair of the gaugino zero modes with those of
        the matter fields exploiting the non-vanishing vacuum value of $\bar\phi^f$,
        Fig. b. Next, we use the same vertex again to trade two remaining gluino zero modes
        into $\psi^2$ (Fig. c). Thus, the one-instanton contribution does generate a
        $\psi\psi/H^4$-term, and supersymmetry ensures that the $F_X/H^3$ term needed for the
        scalar potential is generated as well \cite{affleckdineseibergetal}, although the
        emergence of this term is harder to see (an off-shell background field would be
        required). This contribution also has the $\Lambda^5$-factor needed for dimensional
        reasons.  Higher instanton effects vanish. For instance, two instantons would produce
        $\Lambda^{10}$ in the superpotential, which, as we know on general grounds, is
        impossible.

      \subsubsection{Runaway Vacuum and the Quark Mass Term}
      
        Now that we have established generation  of the superpotential in
        Eq. (\ref{nonperturbativesuperpotential}), we can use the expansion in
        (\ref{oneoverxsquaredexpansion}) and perform the $\d^2\theta$-integration. Note that
        the numerical constant $c$ is fully determined by normalizations of the zero modes,
        which are very well known. For further technical details I refer to \cite{novikov}. 
        
        Eliminating the auxiliary field, we obtain the scalar potential 
        \begin{align}
          V(H)&= \underbrace{V_D}_\text{\clap{$=0$}} + V_F=
          4c\bar{c}\frac{\Lambda^{10}}{\left|H\right|^6}, 
        \end{align}
        which goes to zero at infinity and does not have a minimum at finite field values. Such
        a situation is usually referred to as a {\em runaway vacuum}.
        
        We can, however, cure this disaster by adding a small quark mass to the superpotential
        of the fundamental theory (small for now, we will later take the limit $m\to\infty$, so
        the (s)quarks will drop out again). With the mass term switched on the superpotential
        will take the form 
        \begin{equation}
        W(X) = mX^2 + \frac{c\Lambda^5}{X^2}\,.
        \end{equation}
        Now $\partial W /\partial X$ (and hence $V(H)$) vanishes at finite values of $X$,
        namely, at 
        \begin{align}\label{hsquaredvev}
          H^2&=\pm\sqrt{6c}\,\sqrt{\frac{\Lambda^5}{m}}\gg\Lambda^2\,.
        \end{align}
        These are the minima of the scalar potential. As was expected, there are two
        vacua\,\footnote{A natural question which one might ask is why I quote here the
          solutions for $H^2$ rather than $H$. The point is that $H^2$ is gauge invariant, see
          the definition before Eq. (41).}. The vacuum expectation value of $H^2$ is stabilized
        far away from the origin. This justifies {\em a posteriori} our starting assumption
        that the gauge bosons are very heavy and can be integrated out. 

      \subsubsection{The Konishi Anomaly}
      
        Thus, we found $\langle H^2\rangle$. Remember, however, that our task was to determine
        the gluino condensate, $\langle \lambda^2\rangle$. Since the gluon and gluino fields
        are integrated out, it might seem that information on $\langle \lambda^2\rangle$ is
        lost. This is not the case --- we can relate the vacuum expectation value of $H^2$ to
        the gluino condensate by virtue of  the Konishi anomaly. The Konishi anomaly is a
        supersymmetric generalisation of the chiral anomaly, i.e. the nonconservation of the
        axial current which is proportional to $G_{mn}\widetilde{G}^{mn}$. This product is
        contained in the highest component of the superfield $\tr W^\alpha W_\alpha$, and the
        Konishi relation is
        \begin{align}\label{konishi}
          D^2\left(\Qt Q\right)&=\frac{1}{2\pi}\tr W^\alpha W_\alpha + Q\frac{\partial
            W}{\partial Q},
        \end{align}
        where $W^\alpha$ is the field strength superfield and the $W$ appearing in the last
        term is the superpotential. Spelled out explicitly, the last term 
        is $4m Q^i_f
        Q^j_g\varepsilon^{fg} \varepsilon_{ij}\equiv 4m Q^2$. 
        However, the left hand side of
        Eq. (\ref{konishi}) is a total superspace derivative, so its expectation value in any
        supersymmetric vacuum must vanish. This, in turn, gives us the relation we desire,
        namely
        \begin{align*}
          \left\langle\frac{1}{2\pi}\tr W^\alpha W_\alpha +4m Q^2\right\rangle&=0,
        \end{align*}
        or in particular its lowest component,
        \begin{align}
          \left\langle\frac{1}{2\pi}\lambda^\alpha\lambda_\alpha +4m H^2\right\rangle=0.
        \end{align}
        Inserting the expression for $H^2$ from Eq. (\ref{hsquaredvev}), we get
        \begin{align}\label{lambdalambdakonishi}
          \frac{1}{2\pi}
          \left\langle\lambda^\alpha\lambda_\alpha\right\rangle = \pm \frac{6}{\pi}
          \sqrt{\Lambda^5 m}.
       \end{align}

     \subsubsection{Limit \boldmath$ m\to \infty$: Gluino Condensate, Finally}
     
       The above result was computed at small $m$. However,  in the 
       very end we want to take the limit
       $m\to\infty$. How do we know that Eq. (\ref{lambdalambdakonishi}) is still valid? 
       
       Here
       we can exploit the fact that $\lambda\lambda$ is a chiral quantity which must depend
       {\em analytically} on the (complex) mass parameter $m$, so its functional dependence is
       determined by its singularity structure. Furthermore, we can consider the $R$-symmetry
       of the theory. Invariance of $\d^2\theta\, m X^2$ 
       under the $R$ transformation requires the $R$-charge of $m$ to be
       $-4$. Since the $R$-charge of $\lambda^2$ is $-2$, the analytic $m$ dependence  of
       $\lambda\lambda$ implies that $\lambda\lambda\propto m^k$ with $k=\frac{1}{2}$, which
       means that Eq. (\ref{lambdalambdakonishi}) is exact at all values of $m$. 

       Still, the limit $m\to\infty$ of Eq. (\ref{lambdalambdakonishi}) seems not to make
       sense, since it would imply $\langle\lambda\lambda\rangle\to\infty$ as well. However,
       the scale $\Lambda\equiv\Lambda_\text{1fl}$ is the scale of a theory with one
       flavour. In the $m\to\infty$ limit, the theory reduces again to  pure SYM theory, where
       the relevant scale $\Lambda_\text{SYM}$ may and does differ.  Actually, there is an
       exact result relating the scales,
       \begin{align}
         \Lambda^3_\text{SYM}=\left(\Lambda^5_\text{1fl} m\right)^{\frac{1}{2}},
       \end{align}
       which follows from the matching of the gauge coupling
       running with appropriate $\beta$ functions.
       We see that the right-hand side of this expression is precisely the right-hand side in
       Eq. (\ref{lambdalambdakonishi}). So we can now safely take the limit and arrive at
       \begin{align}
         \left\langle\lambda^\alpha\lambda_\alpha\right\rangle=\pm 12 \Lambda^3_\text{SYM}.
       \end{align}
       This is exactly the expression advertised in Eq. (\ref{gluinocondensate}) restricted to
       $\sf SU(2)$, since the exponential just reduces to $\pm 1$. 
       
     \subsubsection*{Acknowledgements}
       I am very grateful to Misha Shifman for many comments and corrections of the
       manuscript. I also would like to thank the organisers of the Saalburg Summer School for two
       very interesting weeks in the forest.

       \clearpage

  \appendix
  
  \section{Convention and Notation}
    \subsection{Metric}
      We use latin indices $m,n,\dotsc$ for vectors and tensors, the metric and Levi-Civita
      symbols are 
      \begin{align}
        \eta_{mn}&=\diag(+,-,-,-), & \varepsilon_{0123}&=-1.
      \end{align}
      
    \subsection{Weyl Spinors}
      In this notes we are always using two component (Weyl) spinors in van-der-Waerden notation,
      i.e. with dotted and undotted greek letters as indices, indicating the
      $\left(\frac{1}{2},0\right)$ or $(0,\frac{1}{2})$ representation of $\sf
      SL(2,\mathbb{C})$\footnote{This is the universal cover of the Lorentz group.}. These indices
      are raised and lowered with the $\varepsilon$-symbol,  
      \begin{align}
        \chi^\alpha&=\varepsilon^{\alpha\beta}\chi_\beta,
        &\chi_\alpha&=\varepsilon_{\alpha\beta}\chi^\beta,
        &\varepsilon^{12}&=\varepsilon_{21}=1,& 
        \varepsilon^{\alpha\gamma}\varepsilon_{\gamma\beta}&=\delta^\alpha_\beta. 
      \end{align}

      In products of spinors, the convention is that undotted indices are contracted from upper
      left to lower right, and dotted ones from lower right to upper left:
      \begin{align}
        \chi\eta&\equiv\chi^\alpha \eta_\alpha &\chib\etab&\equiv\chib_\alphad\etab^\alphad
      \end{align}
      Since the spinors anticommute and raising and lowering of a contracted index also gives a
      factor of $-1$, we have
      \begin{align}
        \chi\eta&=\chi^\alpha\eta_\alpha=-\eta_\alpha\chi^\alpha=
        -\varepsilon_{\alpha\beta}\eta^\beta\chi^\alpha =\eta^\beta\chi_\beta=\eta\chi
      \end{align}
      und also $\chib\etab=\etab\chib$.

      The r\^{o}le of the $\gamma$ matrices is played by the $\sigma$ matrices,
      \begin{align}
        \sigma^m_{\alpha\alphad}&=\left(-\mathbbm{1}_2,\sigma^i\right)_{\alpha\alphad} \quad
        \text{and} \quad 
        \sigmab^{m\,\alphad\alpha}=\varepsilon^{\alphad\betad}\varepsilon^{\alpha\beta}
        \sigma^m_{\beta\betad}= \left(-\mathbbm{1}_2,-\sigma^i\right)^{\alphad\alpha}
      \end{align}
      with $\sigma^i$ the Pauli matrices. We also need the ``commutator'' of the $\sigma$'s,
      \begin{align}
        \left(\sigma^{mn}\right)_\alpha^{\phantom{\alpha}\beta}=
        \tfrac{1}{4}\left(\sigma^m\sigmab^n-
          \sigma^n\sigmab^m\right)_\alpha^{\phantom{\alpha}\beta} & 
        \left(\sigmab^{mn}\right)^\alphad_{\phantom{\alphad}\betad}=
        \tfrac{1}{4}\left(\sigmab^m\sigma^n-
          \sigmab^n\sigma^m\right)^\alphad_{\phantom{\alphad}\betad}, 
      \end{align}
      which are antisymmetric in $(mn)$. If one of the spinor indices is raised or lowered,
      they are also symmetric in $(\alpha\beta)$ or $(\alphad\betad)$, respectively. 

      Whenever possible, we omit the spinor indices and write e.g. $\chi\eta$ for
      $\chi^\alpha\eta_\alpha$, $\theta^2$ for $\theta^\alpha\theta_\alpha$ and
      $\chi\sigma^m\etab$ for $\chi^\alpha\sigma^m_{\alpha\alphad}\etab^\alphad$.

    \subsection{Relation to Dirac and Majorana Spinors}
      The transition to usual four-component Dirac spinors $\Psi$ is most easily done in the
      chiral (or Weyl) basis for the $\gamma$-matrices,  
      \begin{align}
        \gamma^m&=\begin{pmatrix} 0 &\sigma^m\\\sigmab^m &0\end{pmatrix}.
      \end{align}
      In this basis, a Dirac spinor $\Psi$ directly reduces to two Weyl ones,
      \begin{align}
        \Psi&=\begin{pmatrix} \chi_\alpha \\ \etab^{\alphad}\end{pmatrix}, &
        \Psib&=\left(\eta^\alpha,\chib_{\alphad}\right), 
      \end{align}
      and kinetic and mass terms become
      \begin{align}
        \Psib\gamma^m\partial_m\Psi&=\left(\eta^\alpha,\chib_{\alphad}\right)\begin{pmatrix} 0
          &\sigma^m\\\sigmab^m &0\end{pmatrix}\partial_m\begin{pmatrix} \chi_\alpha \\
          \etab^{\alphad}\end{pmatrix} =\chib\sigma^m\partial_m\chi
        +\eta\sigma^m\partial_m\etab,\\
        m\Psib\Psi&=m\left(\eta^\alpha,\chib_{\alphad}\right)\begin{pmatrix} \chi_\alpha
          \\\etab^{\alphad}\end{pmatrix} = m\left(\eta\chi+\chib\etab\right).
      \end{align}
      
      Majorana spinors $\Psi_M$, on the other hand, correspond to just one Weyl spinor and
      decompose as 
      \begin{align}
        \Psi_M&=\begin{pmatrix} \chi_\alpha \\ \chib^{\alphad}\end{pmatrix}.
      \end{align}

    \subsection{Spinor Derivatives}
      Some care has to be taken when raising ore lowering indices on spinorial
      derivatives (there is an additional minus sign). They act in the following way (note the
      order of the indices on the $\varepsilon$'s):
      \begin{align}
        \frac{\partial}{\partial\theta^\alpha}\theta^\beta&\equiv\partial_\alpha\theta^\beta
        =\delta^\alpha_\beta & \partial_\alpha\theta_\beta&=\varepsilon_{\beta\alpha}
        &\partial^\alpha \theta_\beta&=\delta^\alpha_\beta &\partial^\alpha\theta^\beta
        &=\varepsilon^{\beta\alpha} \\
        \frac{\partial}{\partial\thetab^\alphad}\thetab^\betad
        &\equiv\partialb_\alphad\thetab^\betad =\delta^\alphad_\betad
        &\partialb_\alphad\thetab_\betad&=\varepsilon_{\betad\alphad} & \partialb^\alphad
        \thetab_\betad&=\delta^\alphad_\betad
        &\partialb^\alphad\thetab^\betad&=\varepsilon^{\betad\alphad} 
      \end{align}
      
      Rules for the action of products of $\theta$'s:
      \begin{align}
        \partial_\alpha \theta^\beta\theta_\beta&=\left(\partial_\alpha
          \theta^\beta\right)\theta_\beta -\theta^\beta\partial_\alpha\theta_\beta
        =2\theta_\alpha &\partial^2\theta^2&=\partial^\alpha\partial_\alpha
        \theta^\beta\theta_\beta=2 \partial^\alpha\theta_\alpha=4\\
        \partialb_\alphad \thetab_\betad\thetab^\betad&=\left(\partialb_\alphad
          \thetab_\betad\right)\thetab^\betad -\thetab_\betad\partialb_\alphad\thetab^\betad
        =-2\thetab_\alphad &\partialb^2\thetab^2&=\partialb_\alphad\partialb^\alphad
        \thetab_\betad\thetab^\betad=2 \partialb_\alphad\thetab^\alphad=4
      \end{align}

    \subsection{Vectors and Bispinors}
      Vector indices $m$ can be converted to a pair of spinor ones $\alpha\alphad$ by
      contraction with the $\sigma$ matrices, specifically for vectors $X_m$
      \begin{align}
        X_{\alpha\alphad}&=X_m\sigma^m_{\alpha\alphad} &
        X_m&=-\tfrac{1}{2}X_{\alpha\alphad}\sigmab_m^{\alpha\alphad} & X^m Y_m &=-\tfrac{1}{2}
        X^{\alpha\alphad} Y_{\alpha\alphad},
      \end{align}
      and for antisymmetric tensors $F_{mn}$,
      \begin{align}
        F_\alpha^{\phantom{\alpha}\beta}&=F_{mn}
        \left(\sigma^{mn}\right)_\alpha^{\phantom{\alpha}\beta}
        &F_\alpha^{\phantom{\alpha}\beta}
        \left(\sigma^{mn}\right)_\beta^{\phantom{\beta}\alpha} &= -F^{mn}-\tfrac{1}{2}\i\Ft^{mn},
      \end{align}
      where the dual tensor appears as the imaginary part.
      
    \subsection{Useful Identities}
      Spinor Products:
%%%%%%%%%%%%%%%%%%%%%%%%%%%%%%%%%%%%%%%%%%%%%%%%%%%%%%%%%%%%%%%%%%%%%%%%%%%%%%%%%%%%%%%%%%%%%%%%%%%
      \begin{align}
         \theta^\alpha\theta^\beta & =-\tfrac{1}{2}\varepsilon^{\alpha\beta}\theta\theta,
         &\theta_\alpha\theta_\beta  &=\tfrac{1}{2}\varepsilon_{\alpha\beta}\theta\theta &
         \thetab^{\alphad}\thetab^{\betad}
         &=\tfrac{1}{2}\varepsilon^{\alphad\betad}\thetab\thetab,
         &\thetab_{\alphad}\thetab_{\betad}
         &=-\tfrac{1}{2}\varepsilon_{\alphad\betad} \thetab\thetab\\
         \theta_\alpha \theta^\beta &= -\tfrac{1}{2}\theta \theta \delta_\alpha^\beta,
         &\theta^\alpha \theta_\beta &= \tfrac{1}{2}\theta \theta \delta^\alpha_\beta 
         &\thetab_{\alphad} \thetab^{\betad} &=
         \tfrac{1}{2}\thetab \thetab 
         \delta_{\alphad}^{\betad},
         &\thetab^{\alphad} \thetab_{\betad}
         &= -\tfrac{1}{2}\thetab \thetab \delta^{\alphad}_{\betad} 
       \end{align} 
       
       \noindent Rules for $\sigma$ matrices:
         \begin{align}
           \sigma^m_{\alpha\alphad}\sigmab^{n\,\alphad\beta}&=
           -\eta^{mn}\delta_\alpha^\beta +2(\sigma^{mn})_\alpha^{\phantom{\alpha}\beta} &
           \sigmab^{m\,\alphad\alpha}\sigma^n_{\alpha\betad}&=
           -\eta^{mn}\delta^\alphad_\betad +2(\sigmab^{mn})^\alphad_{\phantom{\alpha}\betad}\\
           (\sigma^{mn})_\alpha^{\phantom{\alpha}\beta}&=
           \tfrac{1}{4}\left(\sigma^m_{\alpha\alphad} 
           \sigmab^{n\,\alphad\beta}-\sigma^n_{\alpha\alphad}\sigmab^{m\alphad\beta}\right)&
           (\sigmab^{mn})^{\alphad}_{\phantom{\alpha}\betad}&
           =\tfrac{1}{4}\left(\sigmab^{m\,\alphad\alpha}\sigma^n_{\alpha\betad}
             -\sigmab^{n\,\alphad\alpha}\sigma^m_{\alpha\betad}\right)\\ 
           (\sigma^{mn})_\alpha^{\phantom{\alpha}\alpha}&=
           (\sigmab^{mn})^{\alphad}_{\phantom{\alpha}\alphad}=0  &
           \tr(\sigma^m\sigmab^n)&=-2\eta^{mn}\\
           \sigma^m_{\alpha\alphad}\sigmab^{\betad\beta}_m &=
           -2\delta_\alpha^\beta\delta_{\alphad}^{\betad} &
           \sigma^m_{\alpha\alphad}\sigma_{m\,\beta\betad}&=
           -2\varepsilon_{\alpha\beta}\varepsilon_{\alphad\betad}
         \end{align}

         \begin{align}
           \sigma^a\sigmab^b\sigma^c &= \eta^{ac}\sigma^b-\eta^{bc}\sigma^a
           -\eta^{ab}\sigma^c+\i\varepsilon^{abcd}\sigma_d\\
           \sigmab^a\sigma^b\sigmab^c &= \eta^{ac}\sigmab^b-\eta^{bc} \sigmab^a
           -\eta^{ab}\sigmab^c-\i\varepsilon^{abcd}\sigmab_d \\
           \sigma^{mn\,\beta}_\alpha \sigma^{kl\,\alpha}_\beta &=
             -\tfrac{1}{2}\left(\eta^{mk}\eta^{nl} -\eta^{ml}\eta^{nk}
               +\i\varepsilon^{mnkl}\right)\\
           \sigmab^{mn\,\betad}_{\phantom{mn\,\betad}\alphad}
           \sigmab^{kl\,\alphad}_{\phantom{kl\,\alphad}\betad} &= 
             -\tfrac{1}{2}\left(\eta^{mk}\eta^{nl} -\eta^{ml}\eta^{nk}
               +\i\varepsilon^{mnkl}\right)
         \end{align}
%%%%%%%%%%%%%%%%%%%%%%%%%%%%%%%%%%%%%%%%%%%%%%%%%%%%%%%%%%%%%%%%%%%%%%%%%%%%%%%%%%%%%%%%%%%%%%%%%%%

    \bibliographystyle{unsrt}

\begin{thebibliography}{99}
      \bibitem{shifmanvainshtein}
        M.~A.~Shifman and A.~I.~Vainshtein,\,``Instantons versus supersymmetry: Fifteen years
        later,'' in M. Shifman {\sl ITEP Lectures on Particle Physics and Field theory} (World
        Scientific, Singapore, 1999), Vol. 2, p. 485, {\tt hep-th/9902018}. 

      \bibitem{shifmanbook}
        M.~A.~Shifman (Ed.), {\sl Instantons in Gauge Theories}
       (World Scientific, Singapore, 1994)

      \bibitem{shifmanvainshteinold}
        M.~A.~Shifman and A.~I.~Vainshtein,
        ``On holomorphic dependence and infrared effects in
        supersymmetric gauge theories,'' Nucl.\ Phys.\ B {\bf 359} (1991) 571. %%CITATION = %%NUPHA,B359,571;%% 

        M.~A.~Shifman and A.~I.~Vainshtein, ``On Gluino Condensation In Supersymmetric Gauge
        Theories. SU(N) And O(N) Groups,'' Nucl.\ Phys.\ B {\bf 296} (1988) 445]. 
%%CITATION = NUPHA,B296,445;%%

 \bibitem{affleckdineseibergetal}
        I.~Affleck, M.~Dine and N.~Seiberg,\,``Dynamical Supersymmetry Breaking In Supersymmetric
        QCD,'' Nucl.\ Phys.\ B {\bf 241}, 493 (1984). 
        %%CITATION = NUPHA,B241,493;%%
        
      \bibitem{wittenindex}
        E.~Witten,\,``Constraints On Supersymmetry Breaking,'' Nucl.\ Phys.\ B {\bf 202}, 253
        (1982). %%CITATION = NUPHA,B202,253;%%
        
\bibitem{golfand}
Y.~A.~Golfand and E.~P.~Likhtman, `` Extension Of The Algebra Of Poincare Group Generators 
 And Violation Of P-Invariance,''
Pisma Zh.\ Eksp.\ Teor.\ Fiz.\  {\bf 13} (1971) 452
[JETP Lett.\  {\bf 13}, 323 (1971); reprinted in {\sl Supersymmetry},
Ed. S. Ferrara, (North-Holland/World Scientific, 1987) Vol. 1, p.  7].
%%CITATION = JTPLA,13,323;%%

\bibitem{haag}
R.~Haag, J.~T.~\textL opuszanski and M.~Sohnius, ``All Possible Generators Of Supersymmetries Of The S Matrix,''
Nucl.\ Phys.\ B {\bf 88}, 257 (1975)
[Reprinted in {\sl Supersymmetry},
Ed. S. Ferrara, (North-Holland/World Scientific, 1987) Vol. 1, p.  51].
%%CITATION = NUPHA,B88,257;%%

\bibitem{salam}
A.~Salam and J.~Strathdee,
``Supergauge Transformations,''
Nucl.\ Phys.\ B {\bf 76}, 477 (1974).
%%CITATION = NUPHA,B76,477;%%

\bibitem{berezin}
F. Berezin, {\sl Methods of Second Quantization},
(Academic Press, New York, 1966).

\bibitem{bailin}
D. Bailin and A. Love, 
{\sl Supersymmetric Gauge Field Theory and String Theory},
(Institute of Physics Publishing, Bristol, 1994). 

\bibitem{novikov}
V.~A.~Novikov, M.~A.~Shifman, A.~I.~Vainshtein and V.~I.~Zakharov,
``Supersymmetric Instanton Calculus: Gauge Theories With Matter,''
Nucl.\ Phys.\ B {\bf 260}, 157 (1985).
%%CITATION = NUPHA,B260,157;%%


\end{thebibliography}

\end{document}